\documentclass[prd,twocolumn,showpacs,nofootinbib,eqsecnum,superscriptaddress,floatfix]{revtex4}
\usepackage{amsmath}
\usepackage{amssymb}
\usepackage{graphicx}
\usepackage{dcolumn}
\usepackage{bm}
\usepackage{verbatim}
\usepackage{hyperref}
\newcommand{\bea}{\begin{eqnarray}}
\newcommand{\eea}{\end{eqnarray}}
\newcommand{\be}{\begin{equation}}
\newcommand{\ee}{\end{equation}}

\newcommand{\et}{\emph{et al.}}
\renewcommand{\case}[2]{\ensuremath{{\textstyle\frac{#1}{#2}}}}

\newcommand{\half}{\ensuremath{{\textstyle\frac{1}{2}}}}

\newcommand{\quarter}{\ensuremath{{\textstyle\frac{1}{4}}}}
\newcommand{\sixth}{\ensuremath{{\textstyle\frac{1}{6}}}}

\newcommand{\Tr}{\mathop{\mathrm{Tr}}}

\newcommand{\ditto}{\multicolumn{1}{c}{~~~"}}

\begin{document}
\title{Quarkonium mass splittings in three-flavor lattice QCD}
\author{T.~Burch}
\affiliation{Department of Physics, University of Utah, Salt Lake City, 
Utah, USA}
\author{C.~DeTar}
\affiliation{Department of Physics, University of Utah, Salt Lake City, 
Utah, USA}
\author{M.~Di~Pierro}
\affiliation{School of Computing, DePaul University, Chicago, Illinois, USA}
\author{A.~X.~El-Khadra}
\affiliation{Physics Department, University of Illinois, Urbana, 
Illinois, USA}
\author{E.~D.~Freeland}
\affiliation{Department of Physics, Washington University, St.~Louis,
Missouri, USA}
\author{Steven~Gottlieb}
\affiliation{Department of Physics, Indiana University, Bloomington, 
Indiana, USA}
\affiliation{National Center for Supercomputing Applications, 
University of Illinois, Urbana, Illinois, USA}
\author{A.~S.~Kronfeld}
\affiliation{Fermi National Accelerator Laboratory, Batavia, Illinois, 
USA}
\author{L.~Levkova}
\affiliation{Department of Physics, University of Utah, Salt Lake City, 
Utah, USA}
\author{P.~B.~Mackenzie}
\affiliation{Fermi National Accelerator Laboratory, Batavia, Illinois, 
USA}
\author{J.~N.~Simone}
\affiliation{Fermi National Accelerator Laboratory, Batavia, Illinois, 
USA}
\collaboration{Fermilab Lattice and MILC Collaborations}
\noaffiliation

\date{\today}
\begin{abstract}
We report on calculations of the charmonium and bottomonium spectrum in 
lattice QCD.
We use ensembles of gauge fields with three flavors of sea quarks,
simulated with the asqtad improved action for staggered fermions.
For the heavy quarks we employ the Fermilab interpretation of the 
clover action for Wilson fermions.
These calculations provide a test of lattice QCD, including the 
theory of discretization errors for heavy quarks.
We provide, therefore, a careful discussion of the results in light of 
the heavy-quark effective Lagrangian.
By and large, we find that the computed results are in agreement with 
experiment, once parametric and discretization errors are taken into 
account.
\end{abstract}

\maketitle

\section{Introduction}
\label{sec:intro}

Quarkonium plays an important role in the application of QCD to hadronic
physics.
Early calculations of charmonium based on potential models gave strong 
support to the interpretation of these states as bound states of a new 
heavy quark~\cite{Quigg:1979vr,Kwong:1987mj}.
Although physically the charmonium state is analogous to positronium,
its historical role as a model system of QCD proved to be analogous to 
that of the hydrogen atom in quantum mechanics.
The charmonium spectrum provided a simple example of how QCD works, 
made even more compelling with the subsequent observation of the 
bottomonium states.

Although the analysis of the quarkonium spectrum based on potential 
models is a significant triumph of QCD, an \emph{ab initio} calculation 
based on lattice QCD, an approach that can simultaneously deal
with light quarks would be even more satisfying.
However, in lattice QCD, the continuum limit requires that $am$ 
approaches zero, where $a$ is the lattice spacing and $m$ is the mass 
of the state.
Quarkonium states and their constituent quarks are so heavy, that it is
often impractical to use so small a lattice spacing that $am$
is small.
Both lattice NRQCD~\cite{Lepage:1987gg,Lepage:1992tx} and the Fermilab 
action~\cite{ElKhadra:1996mp} have been developed to
treat heavy quarks in lattice gauge theory.
Thus, the successful calculation of the spectrum of charmonium and 
bottomonium becomes a significant test of these techniques.

This paper describes the current state of the quarkonium spectrum based
on the Fermilab approach to heavy quarks, using gauge configurations
provided by the MILC Collaboration~\cite{Aubin:2004wf} that incorporate 
the effects of three light quarks: up, down, and strange.
We have been studying the quarkonium spectrum using this formalism
for some time~\cite{DiPierro:2002ta}, starting on ensembles with two 
flavors of sea quark~\cite{ElKhadra:2000zs}, and new
ensembles of configurations have become available during the course 
of the project.
In this paper, we report on results with four lattice spacings from 
$a\approx 0.18$~fm to $\approx 0.09$~fm.
We do not yet consider this work a definitive calculation using our 
approach. 
Results from two finer lattice spacings should become available in 
the future.
As we detail below, the tuning of the bare valence heavy-quark 
masses, via the heavy-light meson spectrum~\cite{Freeland:2009}, 
is not yet precise enough to give satisfactory answers to all the 
questions that have arisen.
In the future, we expect to have better control of the heavy-quark 
masses.
Nevertheless, already we can successfully reproduce important features 
of the quarkonium spectrum, and we consider this another important 
testbed in which to assess the errors that arise from our treatment 
of the heavy quarks.
Knowledge of these errors is also important for calculations of 
properties of heavy-light mesons, for example those pertaining to 
semileptonic decays~\cite{Aubin:2004ej}, 
leptonic decays~\cite{Aubin:2005ar}, 
and $B$-$\bar{B}$ mixing~\cite{Evans:2006zz}, 
as well as the heavy-light spectrum~\cite{Freeland:2009}.

Prior lattice QCD work on quarkonium with different sea-quark content
has been reviewed by Bali~\cite{Bali:2000gf}.
More recently, Dudek and collaborators have used the quenched 
approximation to explore decays~\cite{Dudek:2006ut}, 
radiative transitions~\cite{Dudek:2006ej}, 
and the excited-state spectrum~\cite{Dudek:2007wv} of charmonium.
Meinel has used lattice NRQCD to compute the bottomonium spectrum at 
one lattice spacing on four ensembles with 2+1 domain-wall sea 
quarks~\cite{Meinel:2009rd}.
The HPQCD Collaboration has calculated the quarkonium spectrum on many 
of the same MILC ensembles used in our work, using lattice NRQCD for 
bottomonium~\cite{Davies:2003ik,Gray:2005ur}, and using highly 
improved staggered quarks (HISQ) for charmonium~\cite{Follana:2006rc}.
Lattice NRQCD is not very accurate for charmonium, and HISQ  
requires very small lattice spacings for $b$ quarks~\cite{McNeile:2009eq}.
An advantage of the Fermilab method is that it allows the same 
treatment for both the charmed and bottom quarks.
Predictions of the $c\bar{b}$ spectrum have also been made, for the 
pseudoscalar $B_c$ using NRQCD $b$ quarks and the charmed quark 
propagators from this project~\cite{Allison:2004be}, and also for the 
vector $B_c^*$ using NRQCD $b$ quarks and HISQ charmed 
quarks~\cite{Gregory:2009hq}.

The plan of this paper is as follows.
In Sec.~\ref{sec:methods}, we describe our methodology, explaining the 
actions used for gluons, light sea quarks, and the heavy valence 
quarks.
We describe, in detail, how to use the heavy-quark effective Lagrangian 
to understand heavy-quark discretization errors in quarkonium masses.
We also discuss the construction of the hadronic correlators and how 
they are fit to determine meson masses.
Several broad issues inform the uncertainties (statistical and 
systematic), and they are discussed in Sec.~\ref{sec:issues}.
In Sec.~\ref{sec:results}, we show our results for the splittings 
between various states or combinations of states.
Where possible, we have organized the presentation of the mass 
splittings around individual terms in the heavy-quark effective action, which 
clarifies the approach to the continuum limit at available lattice 
spacings.
Section~\ref{sec:conc} contains our conclusions and suggestions on ways 
to improve on this calculation.

\section{Methodology}
\label{sec:methods}

In this section, we collect several sets of information needed to 
understand the results that follow.
Section~\ref{sec:notation} defines the notation for different 
quarkonium states and splittings.
Then we provide details of the lattice gauge configurations that we 
have used in Sec.~\ref{sec:milc}.
Next, in Sec.~\ref{sec:heavy}, we review the Fermilab method, 
discussing in detail how it can be understood via an effective 
Lagrangian.
This discussion provides a link between the lattice fermion action and 
the computed mass splittings;
an important theme in this paper is to scrutinize our numerical results 
according to these theoretical expectations.
Last, we explain how we form correlation functions in Sec.~\ref{sec:C(t)}, 
and how we fit them to obtain masses in Sec.~\ref{sec:fits}.

\subsection{Notation}
\label{sec:notation}

In this paper, we use two notations for hadrons and their masses, both 
the standard names from the Particle Data Group~\cite{Amsler:2008zz} and 
the spectroscopic notation ${n}^{2S+1}\!L_J$, where $S$, $L$, and $J$ 
are the spin, orbital, and total angular momentum, respectively, of the 
$n$th radial excitation.
As usual, $L=0$, 1, 2, \ldots\ are denoted $S$, $P$, $D$, \ldots.

It is often convenient to discuss spin-averaged masses (and mass 
splittings), which we indicate with a horizontal line, such as 
$\overline{1S}$ or $\overline{1^3\!P}$.
In particular,
\begin{eqnarray}
    M(\overline{1S}) & = & \case{1}{4}\left(M_{\eta_c}+3M_{J/\psi}\right), \\
    M(\overline{1^{3}\!P}) & = & \case{1}{9}\left(
        M_{\chi_{c0}}+3M_{\chi_{c1}}+5M_{\chi_{c2}}\right),
\end{eqnarray}
using charmonium for illustration.
For brevity we usually write $\overline{1P}$ for $\overline{1^3\!P}$.
These spin-averages are sensitive to the leading term in a 
nonrelativistic expansion.
Note that the $\overline{1P}$ and $1^1\!P_1$ (also denoted $h_c$ and $h_b$) 
levels are nearly the same in nature, which can be explained by the 
spin-spin interaction's short range---$\delta(\bm{r})$ in the context of 
potential models~\cite{Kwong:1987mj}.

Complementary to the spin-averaged masses are spin-splittings that hone 
in on spin-dependent corrections~\cite{Eichten:1980mw,Peskin:1983up}.
Below, we examine the hyperfine splittings
\begin{equation}
    M(nS_{\textrm{HFS}}) = M_{J/\psi} - M_{\eta_c},
\end{equation}
using charmonium $1S$ notation on the right-hand side.
For the $P$ states two combinations are of interest
\begin{eqnarray}
    M(nP_{\textrm{spin-orbit}}) & = & \case{1}{9}\left(
        5 M_{\chi_{c2}} - 2 M_{\chi_{c0}} - 3 M_{\chi_{c1}}\right),
    \hspace*{1em} \label{eq:spin-orbit} \\
    M(nP_{\textrm{tensor}})     & = & \case{1}{9}\left(
        3 M_{\chi_{c1}} -   M_{\chi_{c2}} - 2 M_{\chi_{c0}}\right), 
    \label{eq:tensor}
\end{eqnarray}
again using charmonium $1S$ notation on the right-hand side.
$M(nS_{\textrm{HFS}})$ and $M(nP_{\textrm{tensor}})$ are sensitive to 
spin-spin interactions, and $M(nP_{\textrm{spin-orbit}})$ to spin-orbit 
interactions.

\subsection{Configuration details}
\label{sec:milc}

These calculations have been carried out on lattice gauge configurations 
provided by the MILC Collaboration \cite{Aubin:2004wf}, listed in 
Table~\ref{tab:ensembles}.
\begin{table*}[tpb]
\caption[tab:ensembles2]{Run parameters and configuration numbers for
    the ensembles used to study charmonium and bottomonium 
    $\eta$, $J/\psi$, $\Upsilon$, $h$, $\chi_0$, and $\chi_1$ states
    with relativistic operators,
    and $h$ and all $\chi_J$ states with nonrelativistic operators.
    The labels $a$ and $b$ are used to distinguish between the 
    two runs with the same $\beta=6.76$ but different $am_l/am_s$.}
\label{tab:ensembles}
\begin{tabular}{cdr@{/}lc@{\quad\quad}dcdc@{\quad\quad}dcdc}
    \hline\hline
	 & &\multicolumn{2}{c}{ } & & \multicolumn{4}{c}{relativistic} &
	     \multicolumn{4}{c}{nonrelativistic} \\
        $a$~(fm) & \multicolumn{1}{c}{~~~$\beta$} & $am_l$&$am_s$ & 
        $N_s^3\times N_t$ & 
        \multicolumn{1}{c}{~~~$\kappa_c$} & $N_{\textrm{conf}}^c$ & 
        \multicolumn{1}{c}{~~~$\kappa_b$} & $N_{\textrm{conf}}^b$ &
        \multicolumn{1}{c}{~~~$\kappa_c$} & $N_{\textrm{conf}}^c$ & 
        \multicolumn{1}{c}{~~~$\kappa_b$} & $N_{\textrm{conf}}^b$ \\
    \hline
        $\approx 0.18$ & 6.503 & 0.0492&0.082 & $16^3\times48$ &
	    0.120 & 401 & \multicolumn{2}{c}{---} &
	    0.120 & 400 & \multicolumn{2}{c}{---} \\
            & 6.485 & 0.0328&0.082 & " & \ditto & 331 & & & 
            \ditto & 501 & & \\
        & 6.467 & 0.0164&0.082 & " & \ditto & 645 & & & \ditto & 647 & & \\
        & 6.458 & 0.0082&0.082 & " & \ditto & 400 & & & \ditto & 601 & & \\
    \hline
        $\approx 0.15$ & 6.600 & 0.0290&0.0484 & $16^3 \times 48$ & 
	    \multicolumn{2}{c}{---} & \multicolumn{2}{c}{---} &
            0.122 & 580 & 0.076 & 595 \\
	    & 6.586 & 0.0194&0.0484 & " & 0.122 & 631 & 0.076 & 631 &
		    \ditto & 580 & \ditto & 595 \\
        & 6.572 & 0.0097&0.0484 & " & \ditto & 631 & \ditto & 631 &
            \ditto & 629 & \ditto & 631 \\
	    & 6.566 & 0.00484&0.0484 & $20^3 \times 48$ &
	    \multicolumn{2}{c}{---} & \multicolumn{2}{c}{---} &
            \ditto & 601 & \ditto & 600 \\
    \hline
        $\approx 0.12$ & 6.81 & 0.03&0.05 & $20^3\times 64$ &
            0.122 & 549 & 0.086 & 549  &\multicolumn{2}{c}{---} &\multicolumn{2}{c}{---}\\ 
        & 6.79 & 0.02&0.05  & " & \ditto& 460 & \ditto & 460 \\
        & 6.76,a & 0.01&0.05 & " & \ditto& 593 & \ditto & 539 \\
        & 6.76,b & 0.007&0.05 & " & \ditto & 403 & 
            \multicolumn{2}{c}{---}  \\
    \hline
        $\approx 0.09$ & 7.11 & 0.0124&0.031 & $28^3 \times 96$ &
            0.127 & 517 & 0.0923 & 517 & 0.127 & 518 & 0.0923 & 510 \\
        & 7.09 & 0.0062&0.031 & " & \ditto & 557 & \ditto & 557 &
            \ditto & 557 & \ditto & 557 \\
        & 7.08 & 0.0031&0.031 & $40^3 \times 96$ & \ditto & 504 &
            \ditto& 504 &\ditto& 504 &\ditto& 504 \\
    \hline\hline
\end{tabular}
\end{table*}
They were generated via the $R$~algorithm \cite{Gottlieb:1987mq} with 
the one-loop Symanzik-improved L\"uscher-Weisz gluon 
action~\cite{Luscher:1984xn} combined with $2+1$ flavors of sea quarks 
simulated with the asqtad action~\cite{Orginos:1998ue}.
The $n_f$-dependent part of the one-loop couplings~\cite{Hao:2007iz} 
became available only after the ensembles were generated.
We have used ensembles at four lattice spacings: $a\approx0.18$, $0.15$, 
$0.12$, and $0.09$~fm (also called in the text ``extra-coarse'', 
``medium-coarse'', ``coarse'' and ``fine'' ensembles, respectively).
The first four columns of Table~\ref{tab:ensembles} list the parameters
of these ensembles, including the masses of the sea quarks, denoting the 
pair as $am_l/am_s$.
The lattice scale of each ensemble with different sea quark masses was 
kept approximately fixed using the length 
$r_1$~\cite{Sommer:1993ce,Bernard:2000gd} from the static quark 
potential.
The absolute scale from the $\Upsilon$ $2S$-$1S$ splitting was 
determined on most of our ensembles by the HPQCD 
Collaboration~\cite{Davies:2003ik,Gray:2005ur}.
Details on the $r_1$ determinations can be found in review of other 
work on the MILC ensembles~\cite{Bazavov:2009bb}.
Combining this determination with more recent 
work~\cite{Bazavov:2009fk,Davies:2009ts}, leads us to take the range 
$r_1=0.318^{+0.000}_{-0.007}$~fm in this paper.

\subsection{Heavy quark formulation}
\label{sec:heavy}

In this work, the charmed and bottom quarks are simulated with the
Fermilab action~\cite{ElKhadra:1996mp}
\begin{widetext}
\begin{eqnarray}
	S = \sum_n \bar{\psi}_n\psi_n & - & \kappa\sum_n \left[
		\bar{\psi}_n(1-\gamma_4)U_{n,4}\psi_{n+\hat{4}} +
		\bar{\psi}_{n+\hat{4}}(1+\gamma_4)U^\dagger_{n,4}\psi_n 
	\right] \nonumber \\
 	& - & \kappa\zeta \sum_{n,i} \left[
		\bar{\psi}_n(r_s-\gamma_i)U_{n,i}\psi_{n+\hat{\imath}} +
		\bar{\psi}_{n+\hat{\imath}}(r_s+\gamma_i)U_{n,i}^\dagger\psi_n
	\right] \label{eq:S} \\
	& - & c_B\kappa\zeta \sum_n
		\bar{\psi}_ni\bm{\Sigma}\cdot\bm{B}_n\psi_n -
		c_E\kappa\zeta \sum_{n;i} 
		\bar{\psi}_n\bm{\alpha}\cdot\bm{E}_n\psi_n,
	\nonumber
\end{eqnarray}
\end{widetext}
where $U$ denotes the gluon field, 
and $\psi$ and $\bar{\psi}$ denote the quark and antiquark fields.
The clover definitions of the chromomagnetic and chromoelectric 
fields~$\bm{B}$ and~$\bm{E}$ are standard and given, for example, 
in Ref.~\cite{ElKhadra:1996mp}.
When $\zeta=r_s=1$ and $c_B=c_E=0$, $S$ reduces to the Wilson 
action~\cite{Wilson:1977nj}; 
when $\zeta=r_s=1$ and $c_B=c_E=c_{\rm SW}$, it reduces to the 
Sheikholeslami-Wohlert action~\cite{Sheikholeslami:1985ij}.
The relation between the hopping parameter~$\kappa$ and the bare mass is
\begin{equation}
	m_0 a =\frac{1}{2\kappa} - 1 - 3r_s\zeta
	\label{eq:baremass}
\end{equation}
in four space-time dimensions.

To motivate our choices of the input parameters $\kappa$, $\zeta$,
$r_s$, $c_B$, and $c_E$, let us review the nonrelativistic
interpretation of Wilson fermions~\cite{ElKhadra:1996mp}.
The pole energy of a single quark of spatial momentum~$\bm{p}$ is
\begin{equation}
	E(\bm{p}) = m_1 + \frac{\bm{p}^2}{2m_2} + O(p^4),
	\label{eq:energy}
\end{equation}
where the quark rest mass $m_1$ and the kinetic mass~$m_2$ are defined 
at all orders of perturbation theory via the self 
energy~\cite{Mertens:1998wx}.
At the tree level,
\begin{eqnarray}
	m_1 & = & a^{-1}\ln(1+m_0a), \label{eq:m1}  \\
	\frac{1}{m_2} & = & \frac{2\zeta^2}{m_0(2+m_0a)} + 
		\frac{ar_s\zeta}{1+m_0a}. \label{eq:m2}
\end{eqnarray}
In general, $m_1\neq m_2$ unless $m_0a\ll1$; for charmed and bottom
quarks on the ensembles listed in Table~\ref{tab:ensembles} one has 
$m_{0c}a\lesssim1$, $m_{0b}a\gtrsim1$.
One could tune $\zeta$ so that $m_2=m_1$, and we shall revisit that
strategy below.

Equation~(\ref{eq:energy}) is the simplest example of a 
nonrelativistic interpretation of physical quantities computed with 
the action in Eq.~(\ref{eq:S}).
This is justified, because in quarkonium the relative momentum of the 
heavy quarks is small compared with the heavy-quark mass.
This is the basis of the phenomenological success of potential models, 
which yield estimates of the relative velocity and, equivalently, 
internal momentum.
For charmonium
\begin{equation}
	v \approx 0.55, \quad
	p \approx  840~\textrm{MeV},
	\label{eq:ccvp}
\end{equation}
and for bottomonium
\begin{equation}
	v \approx 0.31, \quad
	p \approx 1475~\textrm{MeV}.
	\label{eq:bbvp}
\end{equation}
For both systems the typical kinetic energy is 450~MeV, as seen, for
example, in the 
$\overline{1P}$\/-$\overline{1S}$ splitting.
The kinetic energies $\case{1}{2}m_2v^2$ are small on our lattices, and 
the momenta $m_2v$ are marginally small (especially for bottomonium).

The nonrelativistic interpretation can be extended beyond the tree level
and to higher order in the nonrelativistic expansion using effective
field theories.
This has been pursued in detail emphasizing heavy-light
hadrons~\cite{Kronfeld:2000ck,Harada:2001fi,Harada:2001fj}, and here we
explain the ideas in the context of quarkonium.
As in the Symanzik effective theory, one introduces a continuum
effective Lagrangian, but here it is an effective Lagrangian valid for
heavy quarks.
With quarkonium, the appropriate power-counting rule for the effective 
Lagrangian is that of nonrelativistic QED~\cite{Caswell:1986ui} and
nonrelativistic QCD (NRQCD)~%
\cite{Lepage:1987gg,Lepage:1992tx,Bodwin:1992ui}.
So one has
\begin{equation}
	S \doteq -\sum_s \int d^4x\,\mathcal{L}_{\rm HQ}^{(s)},
	\label{eq:SequivLeff}
\end{equation}
where $s$ counts the powers of velocity.
Here $\doteq$ means that the lattice gauge theory on the left-hand side, 
defined in our case by Eq.~(\ref{eq:S}), is given an effective 
description by the right-hand side.
The first several terms of the effective Lagrangian are
\begin{widetext}
\begin{eqnarray}
	\mathcal{L}_{\rm HQ}^{(2)} & = & 
		-\bar{h}^{(+)}(D_4+m_1)h^{(+)} + 
		\frac{\bar{h}^{(+)}\bm{D}^2h^{(+)}}{2m_2} -
		\bar{h}^{(-)}(D_4+m_1)h^{(-)} + 
		\frac{\bar{h}^{(-)}\bm{D}^2h^{(-)}}{2m_2} ,
	\label{eq:Leff2}  \\
	\mathcal{L}_{\rm HQ}^{(4)} & = & 
		\frac{\bar{h}^{(+)}i\bm{\sigma}\cdot\bm{B}h^{(+)}}{2m_B} +
		\frac{\bar{h}^{(+)}i\bm{\sigma}\cdot(\bm{D}\times\bm{E})h^{(+)}}{8m_E^2} +
		\frac{\bar{h}^{(+)}(\bm{D}\cdot\bm{E})h^{(+)}}{8m_{E'}^2} +
		\frac{\bar{h}^{(+)}(\bm{D}^2)^2h^{(+)}}{8m_4^3} +
		\sixth a^3w_4\bar{h}^{(+)}D_i^4h^{(+)} \nonumber \\
	& + & 
		\frac{\bar{h}^{(-)}i\bm{\sigma}\cdot\bm{B}h^{(-)}}{2m_B} -
		\frac{\bar{h}^{(-)}i\bm{\sigma}\cdot(\bm{D}\times\bm{E})h^{(-)}}{8m_E^2} -
		\frac{\bar{h}^{(-)}(\bm{D}\cdot\bm{E})h^{(-)}}{8m_{E'}^2} +
		\frac{\bar{h}^{(-)}(\bm{D}^2)^2h^{(-)}}{8m_4^3} +
		\sixth a^3w_4\bar{h}^{(-)}D_i^4h^{(-)}, \hspace*{3em} 	
	\label{eq:Leff4}
\end{eqnarray}
\end{widetext}
where $h^{(+)}$ is a two-component field describing the quark,
and   $h^{(-)}$ is a two-component field describing the anti-quark.
The short-distance coefficients $m_1$, $m_2^{-1}$, $m_B^{-1}$, 
$m_E^{-2}$, $m_{E'}^{-2}$, $m_4^{-3}$, and $w_4$ depend on the bare 
quark masses, the bare gauge coupling, and all other couplings of the 
(improved) lattice action.
The terms in $\mathcal{L}_{\rm HQ}^{(s)}$ scale with the heavy quark's
velocity as $v^s$, with the rules~\cite{Lepage:1992tx} 
$\bm{D}\sim m_2v$, $\bm{E}\sim m_2^2v^3$, and $\bm{B}\sim m_2^2v^4$.
In particular, the nonrelativistic kinetic energy, 
$\bm{D}^2/2m_2\sim\half m_2v^2$, is an essential part of quarkonium
dynamics, which is why $m_2$ appears with $v$ in the power counting.
The short-distance coefficients $m_B^{-1}$, $m_E^{-2}$, etc., can be 
expanded in perturbation theory, with $\alpha_s\sim v$ 
\cite{Lepage:1992tx}.
We have put the rest mass $m_1\bar{h}^{(\pm)}h^{(\pm)}$ and
temporal kinetic energy $\bar{h}^{(\pm)}D_4h^{(\pm)}$ into 
$\mathcal{L}_{\rm HQ}^{(2)}$, because by the equation of motion 
$D_4+m_1\sim\bm{D}^2/2m_2\sim\half m_2v^2$.
The next set of terms, $\mathcal{L}_{\rm HQ}^{(6)}$, are not written 
out, because they are numerous yet merely describe subleading 
contributions to the splittings examined below.

One would like to adjust $\kappa$, $\zeta$, $r_s$, $c_B$, and $c_E$ so 
that the lattice gauge theory matches continuum QCD with controllable 
uncertainty.
One would also like to reduce the number of input parameters as much as 
possible, to make the simulation easier to carry out.
The coupling~$r_s$ is redundant: any choice is allowed as long as the 
doubling problem is solved.
We take
\begin{equation}
	r_s = 1.
	\label{eq:rs}
\end{equation}

To derive tuning criteria for the others, one refers to the NRQCD
description of continuum QCD, which takes the same form as
Eqs.~(\ref{eq:Leff2}) and~(\ref{eq:Leff4}), but with the following 
substitutions:
\begin{eqnarray}
	m_1 & \mapsto & m, \label{map:m1} \\
	m_2 & \mapsto & m, \label{map:m2} \\
	\frac{1}{m_B} & \mapsto & \frac{Z_B}{m}, \label{map:mB} \\
	\frac{1}{m_E^2} & \mapsto & \frac{Z_E}{m^2}, \label{map:mE} \\
	\frac{1}{m_{E'}^2} & \mapsto & \frac{Z_{E'}}{m^2}, \label{map:mEp} \\
	\frac{1}{m_4^3} & \mapsto & \frac{Z_4}{m^3}, \label{map:m4} \\
	w_4 & \mapsto & 0,\label{map:w4} 
\end{eqnarray}
where the last is a consequence of Lorentz invariance, as is the 
exact equality of the rest and kinetic masses.
The matching factors $Z_i$ are unity at the tree level and have a 
perturbative expansion.
To bring the lattice field theory in line with continuum QCD, one must
then simply adjust the lattice couplings so that the lattice quantities
on the left in (\ref{map:m1})--(\ref{map:w4}) become, to some accuracy,
the continuum quantities on the right.
In principle, this matching could be carried out 
nonperturbatively~\cite{Lin:2006ur}, although we do not pursue that 
strategy here.

If one restricts one's attention to mass splittings and matrix 
elements, it is not necessary to adjust a coupling to tune~$m_1$.
The operators $\bar{h}^{(\pm)}h^{(\pm)}$ are number operators, 
commuting with everything else in the Hamiltonian~\cite{Kronfeld:2000ck}.
It is therefore acceptable to tolerate a large discretization error in 
the rest mass, and, consequently, one does not need to adjust $\zeta$.
We take
\begin{equation}
	\zeta = 1.
	\label{eq:zeta}
\end{equation}
To obtain the correct dynamics, one must adjust $\kappa$ so that the 
rest of~$\mathcal{L}_{\rm HQ}^{(2)}$ is correctly tuned.
In other words, one must identify the kinetic quark mass~$m_2$ with the
physical quark mass.

The adjustment of $c_B$ stems from a concrete realization 
of~(\ref{map:mB}).
At the tree level
\begin{equation}
	\frac{1}{m_B} = \frac{2\zeta^2}{m_0(2+m_0a)} + 
		\frac{ac_B\zeta}{1+m_0a}, 
	\label{eq:mB}
\end{equation}
so to ensure $m_B=m_2$ (as desired at the tree level where $Z_B=1$), 
one needs $c_B=r_s$.
In practice, we take [recalling Eq.~(\ref{eq:rs})]
\begin{equation}
	c_B = u_0^{-3}
	\label{eq:cBu0}
\end{equation}
to account for tadpole diagrams at higher orders in perturbation 
theory~\cite{Lepage:1992xa}.
On the coarse ensembles, we set $u_0$ from the Landau link;
on the other ensembles, we set it from the plaquette.

In principle, the adjustment of $c_E$ should stem from~(\ref{map:mE}).
These simulations have been carried out, however, in concert with
calculations of heavy-light masses~\cite{Freeland:2009}, for which the 
adjustment of $c_E$ is a subleading 
effect~\cite{Kronfeld:2000ck,Christ:2006us}.
Thus, we have taken
\begin{equation}
	c_E = c_B.
	\label{eq:cE}
\end{equation}
Using formulae in Ref.~\cite{Oktay:2008ex}, we can estimate the error
stemming from $1/m_E^2$, finding a tree-level mismatch of
\begin{equation}
	\frac{1}{4m_E^2}-\frac{1}{4m_2^2} =
		\frac{a^2}{(2+m_0a)(1+m_0a)} - \frac{a^2}{4(1+m_0a)^2},
	\label{eq:mEm2}
\end{equation}
where the right-hand side holds for $\zeta=r_s=c_B=c_E=1$.
At the tree level $m_E=m_{E'}$, so the same error is made in the 
Darwin terms $\bar{h}^{(\pm)}\bm{D}\cdot\bm{E}h^{(\pm)}$.

An advantage of using Eqs.~(\ref{eq:SequivLeff}), (\ref{eq:Leff2}) 
and (\ref{eq:Leff4}) to describe our lattice calculation is that it 
clarifies which parameters in $S$ play a key role in various splittings 
defined in Sec.~\ref{sec:notation}.
The spin-averaged masses receive energy (beyond $2m_1$) from the balance 
between the kinetic energies $\bar{h}^{(\pm)}\bm{D}^2h^{(\pm)}$ and 
the exchange of temporal gluons between $\bar{h}^{(+)}A_4h^{(+)}$ and
$\bar{h}^{(-)}A_4h^{(-)}$.
As discussed above, they are sensitive to $m_2$, motivating the tuning 
of $\kappa$ (and the fixed choice for $\zeta$.)
The hyperfine splittings $M(nS_{\textrm{HFS}})$ arise from exchange 
of spatial gluons between $\bar{h}^{(+)}i\bm{\sigma}\cdot\bm{B}h^{(+)}$ 
and $\bar{h}^{(-)}i\bm{\sigma}\cdot\bm{B}h^{(-)}$.
Hence they are proportional to~$1/m^2_B$ and, drilling further back 
to~$S$, sensitive to the coupling~$c_B$.
The same line of dependency holds for the tensor splittings 
$M(nP_{\textrm{tensor}})$.
Similarly, the spin-orbit part of the $\chi_{cJ}$ and $\chi_{bJ}$ levels 
arise from exchange of a temporal gluon between
$\bar{h}^{(\pm)}i\bm{\sigma}\cdot(\bm{D}\times\bm{E})h^{(\pm)}$ and 
$\bar{h}^{(\mp)}A_4h^{(\mp)}$.
Hence they are proportional to $1/m_E^2$ and, referring back to $S$, 
sensitive to~$c_E$.

With the tree-level adjustment of $c_B$, the hyperfine splittings 
should be expected to have errors of order $\alpha_s mv^4$ from 
radiative corrections to $m_B^{-1}$
[relative error: ${\rm O}(\alpha_s)\sim{\rm O}(v)$],
and of order $v^6$ from the terms 
$\bar{h}^{(\pm)}\{\bm{D}^2,i\bm{\sigma}\cdot\bm{B}\}h^{(\pm)}$ in
$\mathcal{L}_{\rm HQ}^{(6)}$ [relative error: ${\rm O}(v^2)$].
Similarly, with $c_E=c_B$, we expect leading errors of order $a^2m^3v^4$ 
in the spin-orbit part of the $\chi$ splittings 
[relative error: ${\rm O}(m^2a^2)$], as well as radiative corrections 
to~$m_E^{-2}$ [relative error: again ${\rm O}(\alpha_s)\sim{\rm O}(v)$].
On the MILC ensembles both relative errors are expected to be a few to 
several percent~\cite{Oktay:2008ex}, and, perhaps counterintuitively, 
smaller for bottomonium than charmonium~\cite{Oktay:2008ex}.

The lattice action in Eq.~(\ref{eq:S}) does not contain parameters to 
tune the two terms proportional to $p^4$ in Eq.~(\ref{eq:Leff4}).
The mismatches 
\begin{widetext}
\begin{equation}
    \frac{1}{8m_4^3} -\frac{1}{8m_2^3} = 
        \frac{a^2}{2m_0(2+m_0a)^2(1+m_0a)} +
        \frac{a^2(1+4m_0a)}{4m_0(2+m_0a)(1+m_0a)^2} +
        \frac{m_0a^4}{8(1+m_0a)^3} 
    \label{diff:m4m2}
\end{equation}
and
\end{widetext}
\begin{equation}
	a^3w_4 = \frac{2a^2}{m_0(2+m_0a)} + \frac{a^3}{4(1+m_0a)}
    \label{diff:w4}
\end{equation}
(given again for $\zeta=r_s=c_B=c_E=1$)
cause errors of order $a^2m^3v^4$ in the spin-averaged splittings.
The relative errors, ${\rm O}(m^2a^2)$, are again expected to be a 
few to several percent, but in this case larger for bottomonium than 
for charmonium~\cite{Oktay:2008ex}.
For plots of the $a$ dependence of discretization effects caused by 
Eqs.~(\ref{eq:mEm2}), (\ref{diff:m4m2}), and~(\ref{diff:w4}), 
see Figs.~2 and~3 of Ref.~\cite{Oktay:2008ex}.

To tune $\kappa$ nonperturbatively, one adjusts it so that a hadron mass 
agrees with the experimentally measured value.
Let us define $M_1$ and $M_2$ for a hadron analogously to 
Eq.~(\ref{eq:energy}).%
\footnote{In this paper, we use $m_1$, $m_2$, \ldots\ for quark masses,
and $M_1$, $M_2$, \ldots\ for hadron masses.}
From the effective Lagrangian description for quarkonium,
Eqs.~(\ref{eq:SequivLeff})--(\ref{eq:Leff4}), it follows that
\begin{eqnarray}
    M_1 & = & 2m_1 + B_1, \\
    M_2 & = & 2m_2 + B_2,
\end{eqnarray}
where the binding energy $B_1$ is determined by terms of order $v^2$ 
and higher, but $B_2$ by terms of order $v^4$ and 
higher~\cite{Kronfeld:1996uy}.
In the splittings of rest masses, $m_1$ drops out, so we can obtain 
well-tuned results for $B_1$ (and their differences) by adjusting 
$\kappa$ so that $m_2$ corresponds to a physical quark.
That suggests tuning $\kappa$ so that, say, $M_2(\overline{1S})$ agrees 
with experiment.
The spin average is useful, because it eliminates the leading effect of 
a mistuned chromomagnetic coupling~$c_B$.

A better approach, still using a hadron's kinetic mass, is as follows.
Reference~\cite{Kronfeld:1996uy} analyzes the Breit equation to show 
how higher-order potentials and the $p^4$ terms generate $B_2$, 
tracing how the mismatches noted in Eqs.~(\ref{diff:m4m2}) 
and~(\ref{diff:w4}) propagate to~$B_2$.
This analysis reveals that the discretization error in~$B_2$ is 
smaller for heavy-light hadrons than for quarkonium states.
For heavy-light hadrons, the largest part of the kinetic binding energy
comes from the light quarks and gluons, and, since the light quark has
mass $ma\ll 1$, its contribution to the kinetic binding energy of the
meson has only a small discretization error.
To tune $\kappa$ for charmed and bottom quarks, it is therefore better 
to use heavy-light states, such as $D_s^{(*)}$ and $B_s^{(*)}$, whose 
kinetic masses have the smallest statistical, discretization, and chiral 
extrapolation errors.
In fact, the leading discretization error, from the chromomagnetic 
energy, can again be removed by taking the spin-averaged mass of the 
pseudoscalar and vector mesons.

It is sometimes thought that the tuning inaccuracy of the kinetic 
binding energy~$B_2$ can be circumvented by adjusting $\zeta$ so that 
(a hadron's) $M_1=M_2$, and then fixing $M_1$ to experiment.
But any discretization error in $B_2$ is then propagated to $\zeta$ 
and, hence, throughout the rest of the simulation.
It is, therefore, just as clean to leave $\zeta=1$ and tune $M_2$ to a 
target meson mass, as we have done here.

At this stage, it may be helpful to compare and contrast the Fermilab 
approach~\cite{ElKhadra:1996mp,Oktay:2008ex} with 
lattice NRQCD~\cite{Lepage:1987gg,Lepage:1992tx}.
The construction of lattice NRQCD starts with the (dimensionally 
regulated and $\overline{\rm MS}$-renormalized) NRQCD effective 
Lagrangian for continuum QCD~\cite{Caswell:1986ui,Bodwin:1992ui}, 
and then discretizes it.
This process can be repeated order-by-order in perturbation theory.
In the Fermilab method, a version of the Wilson-Sheikholeslami-Wohlert 
lattice action is used, but the results are interpreted with 
(dimensionally regulated and $\overline{\rm MS}$-renormalized) NRQCD 
with modified short-distance coefficients.
This is possible because Wilson fermions possess heavy-quark symmetry,
and the proposed improvements preserve this feature.
Then the parallel structure of the NRQCD descriptions of QCD and 
lattice gauge theory are used to match the latter to the former.
In both frameworks, the lattice action can be systematically improved 
via the nonrelativistic expansion~\cite{Lepage:1992tx,Oktay:2008ex}.

At a practical level, early spectrum calculations~\cite{Davies:1994mp}
use a lattice-NRQCD action~\cite{Lepage:1992tx} that adjusts, at the 
tree level, the full $v^4$ Lagrangian and the spin-dependent $v^6$ 
Lagrangian.%
\footnote{The HPQCD Collaboration's most recently published unquenched 
calculations~\cite{Gray:2005ur} of the bottomonium spectrum with lattice 
NRQCD are obtained from an action without the spin-dependent $v^6$ 
corrections.}
The $p^4$ terms are, thus, correctly normalized at the tree level, so 
the quarkonium and heavy-light kinetic mass tunings are comparably 
accurate.
On the other hand, the Fermilab action has tree-level errors in the
$v^6$ and even some of the $v^4$ terms.
The errors diminish monotonically as $a$ is reduced, however.
This is especially important for charmonium: here the nonrelativistic 
expansion is not especially good, but it is needed only to organize the 
matching of the most important couplings in $S$, knowing that further 
errors, such as those described by $\mathcal{L}_{\rm HQ}^{(6)}$, 
are of the form $(mv^2a)^2$ and smaller.

In summary, the pattern of discretization effects leads us to tune 
$\kappa$ via kinetic masses corresponding to the 
$\overline{1S}$ $D_s$ and $B_s$ mesons.
The main spectroscopic results, presented in Sec.~\ref{sec:results}, are 
for mass splittings, in which case the uncertainties are minimized by 
quoting differences of our computed rest masses.

\subsection{Correlator construction}
\label{sec:C(t)}

The meson correlator at a given spatial momentum $\bm{p}$ and
time $t$ is defined as
\be
    C_{ab}(\bm{p},t) = \sum_{\bm{x}}e^{-i\bm{p}\cdot\bm{x}}
        \langle 0|O_a(\bm{x},t)O_b^{\dag}(\bm{0},0)|0\rangle,
\ee
where $\bm{x}$ is the spatial coordinate.
The source and sink meson operators $O_{b}$ and $O_a$ have the form
\be
    O_c(\bm{x},t) = \sum_{\bm{y}}
        \bar{\psi}(\bm{x},t)\Gamma\phi_c(\bm{x}-\bm{y})\psi(\bm{y},t),
\ee
where $\Gamma$ is a product of Dirac matrices appropriate for the meson 
spin structure, and $\phi_c(\bm{x}-\bm{y})$ is a smearing function.
Neglecting the disconnected piece, the meson correlator can be 
re-written with the quark propagators 
\be
    G(\bm{x},t;\bm{0},0) = \int [d\psi][d\bar{\psi}]
        \psi(\bm{x},t)\bar{\psi}(\bm{0},0) e^{-S},
\ee
with $S$ from Eq.~(\ref{eq:S}), yielding
\bea
    C_{ab}(\bm{p},t) & = & \sum_{\bm{x}}e^{-i\bm{p}\cdot\bm{x}}
        \times \\ \nonumber
    & & \hspace{2em}
        \Tr\left[G(\bm{0},0;\bm{x},t)\Gamma G_{ab}(\bm{x},t;\bm{0},0)
            \Gamma^\dag\right],
\eea
where
\be
    G_{ab}(\bm{x},t;\bm{0},0)=\sum_{\bm{y},\bm{z}}
        \phi_a(\bm{x}-\bm{y})G(\bm{y},t;\bm{z},0)\phi_b^\dag(\bm{z})
\ee
is the smeared quark propagator.

For the $P$ states, we use two types of quarkonium correlators, which 
we call ``relativistic'' and ``nonrelativistic.''
In the relativistic case, all four spin components
of the quark propagators were used to construct the two-point functions. 
We used point and smeared sources and sinks.
The smearing functions $\phi_c(\bm{x})$ are $1S$ and $2S$ wavefunctions 
of the QCD-motivated Richardson potential~\cite{Richardson:1978bt}.
At the sink, spatial momentum $\bm{p}=2\pi(n_1,n_2,n_3)/L$ is given to 
the quarkonium state.
We restrict the range of $\bm{p}$ such that $\sum n^2_i\leq 9$.
Using this approach,
we computed correlation functions for the $1S$ and $2S$ states for the
pseudoscalar and the vector to study both the kinetic and rest masses.
For the $1P$ states $h$, $\chi_0$ and~$\chi_1$
we computed only the rest masses.

In the nonrelativistic approach to constructing the two-point functions,
the meson operators project onto two of the Dirac components 
of the quark fields.
Table~\ref{tab:pw_op} gives the explicit form of these operators.
\begin{table}
\centering
    \caption[tab:pw_op]{Nonrelativistic meson operators for the $1P$ 
        states.
        The smearing operator in spatial direction $i$ is denoted by~$p_i$.
        The indices $j$ and $k$ are different from $i$ and each other, and 
        repeated indices on the last line are not summed over.}
    \label{tab:pw_op}
    \begin{tabular}{cccll}
        \hline\hline
        Meson & ${{}^{2S+1}\!L_J}$ & Irrep. & \multicolumn{2}{c}{Operator} \\
        \hline
        $h$   & $^1\!P_1$ & $T_1$  & $p_i$, & $i=1,2,3$   \\
     $\chi_0$ & $^3\!P_0$ & $A_1$  & $\sum_{i=1}^3 \sigma_ip_i$ & \\
     $\chi_1$ & $^3\!P_1$ & $T_1$  & $\sigma_j \times p_k$, & $i=1,2,3$ \\
     $\chi_2$ & $^3\!P_2$ & $T_2$  & $\sigma_j p_k + \sigma_k p_j$, & $i=1,2,3$ \\
     $\chi_2$ & $^3\!P_2$ & $E_2$  & $\sigma_j p_j - \sigma_k p_k$, & $i=1,2$ \\
        \hline\hline
    \end{tabular}
\end{table}
At the source and sink we smear the quark propagators with a $P$-type
wavefunction $\phi_c(\bm{r}) = \phi_{1S}(|\bm{r}|)\hat{r}_i$ where
$\phi_{1S}(|\bm{r}|)$ is a Richardson 1S 
wavefunction~\cite{Richardson:1978bt} and $i = 1,2,3$.
At the origin we set $\phi_c(\bm{0})=0$.
The relativistic interpolating operators include extra lower Dirac 
components that increase the overlap with excited states.
Therefore, one should expect that the overlap of the nonrelativistic 
meson operators with the $1P$ ground states to be better than in the 
relativistic case.
We used these nonrelativistic operators at $\bm{p}=\bm{0}$ for the 
$h$, $\chi_0$, $\chi_1$, and $\chi_2$ states.
In Sec.~\ref{sec:relnonrel}, we compare the results for the first three 
states with the corresponding results from relativistic operators.

For both correlator constructions, we use several time-slice positions 
for the source vectors.
In the case of the coarse $\beta=6.76$, $am_l/am_s=0.005/0.05$ ensemble 
and all medium-coarse ensembles, we use eight sources for the 
relativistic operators; in all other cases, we use four.

\subsection{Fitting methods}
\label{sec:fits}

To determine the mass spectrum, we fit our correlator data with a
Bayesian procedure, taking priors guided by
potential models \cite{Richardson:1978bt,Buchmuller:1980su}. 
The priors, listed in Table~\ref{tab:priors}, are the same for both 
relativistic and nonrelativistic correlators.
\begin{table}[tpb]
\centering
\caption{Prior central values for the ground-state masses.
    The priors' widths are all fixed to 0.5.}
\label{tab:priors}
\begin{tabular}{cdd}
    \hline\hline
        $a$~(fm) & \multicolumn{1}{c}{~~$\kappa$} & 
        \multicolumn{1}{r}{$M_{q\bar{q}}a$~~~} \\
    \hline
        $\approx 0.18$ & 0.120 & 1.932386 \\
    \hline
        $\approx 0.15$ & 0.122 & 1.841549 \\
            & 0.076 & 3.818718 \\
    \hline
        $\approx 0.12$ & 0.122 & 1.539279 \\
            & 0.086 & 3.187431 \\
    \hline
        $\approx 0.09$ & 0.127 & 1.184840 \\
            & 0.0923 & 2.818421 \\
    \hline\hline
\end{tabular}
\end{table}
To find the quarkonium masses from relativistic correlators, we use a 
delta function and a $1S$ smearing wavefunction as the source and sink.
We fit simultaneously two or three source-sink combinations for the 
zero-momentum states, including the ground state and up to two excited 
states.
The minimum and maximum source-sink separation is varied, and the best 
fit is selected based on the confidence level and the size of the 
errors in the ground state and first excited state masses.
After choosing the fit range, 250 bootstrap samples are generated to 
provide an error estimate.

The fitting method in the case of nonrelativistic operators is similar 
except we use the same $P$-type wavefunction, described above, for both 
source and sink.
In this case, we use no more than a ground state plus one excited state 
in the fitting form.
The quality of data in the nonrelativistic case is such that often a fit 
with just the ground state is enough, provided the fitting range is 
appropriately chosen.

\section{General Results}
\label{sec:issues}

Before presenting results for mass splittings (in 
Sec.~\ref{sec:results}) we discuss three general issues:
a comparison of the statistical quality of relativistic and 
nonrelativistic operators (Sec.~\ref{sec:relnonrel});
a numerical comparison of tuning $\kappa$ via $M_2$ in heavy-light 
and quarkonium (Sec.~\ref{sec:Qvsq});
and a discussion of how uncertainties from tuning $\kappa$ are 
propagated to the mass splitting (Sec.~\ref{sec:kappa}). 
%
%
\begin{table*}
\centering
\caption{Rest masses of the charmonium states $\eta_c$, $J/\psi$, $h_c$ 
    $\chi_{c0}$, and $\chi_{c1}$ calculated with relativistic operators.
    All masses in units of $r_1=0.318$~fm.
    The star denotes masses that differ from their counterparts in 
    Table~\ref{tab:cc-nonrel} by more than 1.5$\sigma$.}
\label{tab:cc-rel}
\begin{tabular*}{\textwidth}{c@{\extracolsep{\fill}}l%
@{\extracolsep{\fill}}c*{4}{@{\extracolsep{\fill}}l}%
*{3}{@{\extracolsep{\fill}}r@{\hspace*{-1.5em}}l}}
    \hline\hline
$a$~(fm) & \multicolumn{1}{c}{$\beta$} & $\kappa_c$ & $\eta_c(1{}^1S_0)$ &
    $\eta_c(2{}^1S_0)$ & $J/\psi(1{}^3S_1)$ & $\psi(2{}^3S_1)$ &
    & \multicolumn{1}{c}{$h_c(1{}^1P_1)$} & 
    & \multicolumn{1}{c}{$\chi_{c0}(1{}^3P_0)$} &
    & \multicolumn{1}{c}{$\chi_{c1}(1{}^3P_1)$} \\
\hline
$\approx 0.18$ & 6.503 & 0.120 
&3.2924(9)
&4.24(6)
&3.4452(16)
&4.35(7)
&       & 4.185(17)
&$\star$& 4.079(12)
&$\star$& 4.052(89)\\
&6.485 & "
&3.3071(14)
&4.42(4)
&3.4581(18)
&4.48(3)
&       & 4.214(26)
&       & 4.117(15)
&$\star$& 4.173(13)\\
&6.467 & " 
&3.3327(7)
&4.39(27)
&3.4862(11)
&4.45(11)
&       & 4.213(29)
&       & 4.109(12)
&       & 4.200(25)\\
&6.458& "
&3.3481(13)
&4.47(6)
&3.5004(16)
&4.43(10)
&$\star$& 4.217(18)
&       & 4.106(18)
&$\star$& 4.181(19)\\
\hline
$\approx 0.15$ & 6.586 & 0.122
&3.5688(8)
&4.66(3)
&3.7317(13)
&4.75(3)
&$\star$& 4.476(8)
&       & 4.341(6)
&       & 4.450(7)\\
&6.572& "
&3.5883(9)
&4.64(5)
&3.7501(14)
&4.79(2)
&$\star$&4.495(8)
&       &4.368(6)
&       &4.471(17) \\
\hline
$\approx 0.12$ & 6.81 & 0.122
&3.8721(11) 
&5.16(4) 
&4.0594(18) 
&5.25(3)
&       & 4.807(17) 
&       & 4.626(10) 
&       & 4.755(19) \\
&6.79 &"
&3.8876(12) 
&5.14(3) 
&4.0747(18)  
&5.22(4)  
&       & 4.821(12) 
&       & 4.657(10)
&       & 4.791(10)\\
&$6.76,a$&"
&3.8824(9)
&5.09(4)
&4.0677(15)
&5.10(5)
&       &  4.800(13)
&       & 4.658(8) 
&       & 4.758(15)\\
&$6.76,b$&"
&3.9009(8)
&5.12(3)
&4.0864(11)
&5.27(3)
&       &4.817(14) 
&       & 4.650(30) 
&       &4.785(15) \\
\hline
$\approx 0.09$&7.11&0.127
&4.2740(26)
&5.33(13)
&4.4460(22)
&5.55(6)
&       & 5.159(29)
&       & 5.027(16)
&       & 5.185(10)\\
&7.09& "
&4.2885(15)
&5.52(4)
&4.4596(15)
&5.66(4)
&$\star$& 5.149(24)
&$\star$& 4.986(15)
&$\star$& 5.123(19)\\
&7.08& "
&4.2889(26)
&5.51(5)
&4.4613(33)
&5.65(7)
&       & 5.167(33)
&$\star$& 4.986(26)
&       & 5.133(48)\\
\hline\hline
\end{tabular*}
\end{table*}
\begin{table*}
\centering
\caption{Rest masses of the charmonium states $h_c$, $\chi_{c0}$, 
    $\chi_{c1}$, and $\chi_{c2}$ calculated with nonrelativistic 
    operators.
    All masses in units of $r_1=0.318$~fm.
    The star denotes masses that differ from their counterparts in 
    Table~\ref{tab:cc-rel} by more than 1.5$\sigma$.}
\label{tab:cc-nonrel}
\begin{tabular*}{0.75\textwidth}{c@{\extracolsep{\fill}}l%
@{\extracolsep{\fill}}c*{3}{@{\extracolsep{\fill}}r@{\hspace*{-2em}}l}l}
        \hline\hline
$a$~(fm) & \multicolumn{1}{c}{$\beta$} & $\kappa_c$ & 
    & \multicolumn{1}{c}{$h_c(1{}^1P_1)$} & 
    & \multicolumn{1}{c}{$\chi_{c0}(1{}^3P_0)$} & 
    & \multicolumn{1}{c}{$\chi_{c1}(1{}^3P_1)$} & 
      \multicolumn{1}{c}{$\chi_{c2}(1{}^3P_2)$} \\
\hline
$\approx 0.18$ & 6.503 & 0.120 
&       & 4.213(1)
&$\star$& 4.111(9)
&$\star$& 4.210(9)
& 4.272(15)\\
&6.485& "
&       & 4.227(7)
&       & 4.105(8)
&$\star$& 4.200(7)
& 4.286(10)\\
&6.467& "
&       & 4.223(12)
&       & 4.127(7)
&       & 4.227(8)
& 4.278(15)\\
&6.458& "
&$\star$& 4.253(9)
&       & 4.128(9)
&$\star$& 4.222(9)
& 4.310(11)\\
\hline
$\approx 0.15$&6.600&0.122 
&       & 4.492(7)
&       & 4.344(6)
&       & 4.458(7)
& 4.537(11)\\
&6.586& "
&$\star$& 4.493(7)
&       & 4.349(7)
&       & 4.462(7)
& 4.536(13)\\
&6.572& "
&$\star$& 4.516(9)
&       & 4.375(9)
&       & 4.488(9)
& 4.574(10)\\
&6.566& "
&       & 4.548(10)
&       & 4.405(6)
&       & 4.526(7)
& 4.614(10)\\
\hline
$\approx 0.09$&7.11&0.127
&       & 5.199(11)
&       & 5.030(8)
&       & 5.170(12)
& 5.257(12)\\
&7.09& "
&$\star$& 5.198(13)
&$\star$& 5.034(11)
&$\star$& 5.168(13)
& 5.257(14)\\
&7.08 & "
&       & 5.178(15)
&$\star$& 5.047(8)
&       & 5.167(13)
& 5.232(18)\\
\hline\hline
\end{tabular*}
\end{table*}
%
%
\begin{table*}
\centering
\caption{Rest masses of the bottomonium states $\eta_b$, $\Upsilon$, 
    $h_b$, $\chi_{b0}$, and $\chi_{b1}$ calculated with relativistic 
    operators.
    All masses in units of $r_1=0.318$~fm.
    The star denotes masses that differ from their counterparts in 
    Table~\ref{tab:bb-nonrel} by more than 1.5$\sigma$.}
\label{tab:bb-rel}
\begin{tabular*}{\textwidth}{c@{\quad}l@{\quad}c*{4}{@{\hspace*{-1.5em}}d}%
@{\extracolsep{\fill}}r@{\hspace*{-4em}}d%
@{\extracolsep{\fill}}r@{\hspace*{-3.5em}}d%
@{\extracolsep{\fill}}r@{\hspace*{-3.5em}}d}
    \hline\hline
$a$~(fm) & \multicolumn{1}{d}{\beta} & $\kappa_b$ & 
    \multicolumn{1}{r}{$\eta_b(1{}^1S_0)\hspace*{-1em}$} &
    \multicolumn{1}{r}{$\eta_b(2{}^1S_0)\hspace*{-1.2em}$} & 
    \multicolumn{1}{r}{$\Upsilon(1{}^3S_1)\hspace*{-1em}$} & 
    \multicolumn{1}{r}{$\Upsilon(2{}^3S_1)$} &
    & \multicolumn{1}{r}{$h_b(1{}^1P_1)\hspace*{0.2em}$} & 
    & \multicolumn{1}{r}{$\chi_{b0}(1{}^3P_0)$} &
    & \multicolumn{1}{r}{$\chi_{b1}(1{}^3P_1)$} \\
\hline
$\approx 0.15$&6.586&0.076
&7.3776(8)
&8.202(5)
&7.4100(9)
&8.209(5)
&       & 8.269(120)
&$\star$& 8.162(35)
&$\star$& 8.147(40)\\
&6.572&"
&7.4061(9)
&8.241(63)
&7.4386(9)
&8.248(7)
&       & 8.321(13)
&       & 8.292(11)
&       & 8.318(11)\\
\hline
$\approx 0.12$&6.81&0.086
&8.0690(10)
&8.933(12)
&8.1299(12)
&8.957(12)
&       &8.919(15)
&       &8.855(13)
&       &8.898(13)\\
&6.79&"
&8.0563(17)
&8.910(14)
&8.1167(19)
&8.929(14)
&  & 8.902(19)
&  &8.850(21)
&  &8.874(38)\\
&$6.76,a$&"
&7.9815(9)
&8.870(11)
&8.0426(13)
&8.890(10)
&& 8.860(20)
&& 8.796(13)
&& 8.839(14)\\
\hline
$\approx 0.09$&7.11&0.0923
&10.2040(15)
&11.130(78)
&10.2627(19)
&11.160(32)
&       & 11.050(14)
&$\star$& 11.006(13)
&$\star$& 11.049(10)\\
&7.09&"
&10.1861(11)
&11.142(20)
&10.2468(15)
&11.161(18)
&$\star$& 11.056(19)
&       & 10.992(14)
&       & 11.034(13)\\
&7.08&"
&10.1795(26)
&11.112(30)
&10.2397(33)
&11.137(56)
&$\star$& 11.066(19)
&$\star$& 11.017(12)
&$\star$& 11.048(14)\\
\hline\hline
\end{tabular*}
\end{table*}
\begin{table*}
\centering
\caption{Rest masses of the charmonium states $h_b$, $\chi_{b0}$, 
    $\chi_{b1}$, and $\chi_{b2}$ calculated with nonrelativistic 
    operators.
    All masses in units of $r_1=0.318$~fm.
    The star denotes masses that differ from their counterparts in 
    Table~\ref{tab:bb-rel} by more than 1.5$\sigma$.}
\label{tab:bb-nonrel}
\begin{tabular*}{0.75\textwidth}{c@{\extracolsep{\fill}}l%
@{\extracolsep{\fill}}c*{3}{@{\extracolsep{\fill}}r@{\hspace*{-4.5em}}d}%
@{\hspace*{-2em}}d}
        \hline\hline
$a$~(fm) & \multicolumn{1}{c}{$\beta$} & $\kappa_b$ & 
    & \multicolumn{1}{r}{$h_b(1{}^1P_1)$} & 
    & \multicolumn{1}{r}{$\chi_{b0}(1{}^3P_0)$} & 
    & \multicolumn{1}{r}{$\chi_{b1}(1{}^3P_1)\hspace*{-2em}$} & 
      \multicolumn{1}{r}{$\chi_{b2}(1{}^3P_2)$} \\
\hline
$\approx 0.15$&6.600&0.076
&       & 8.254(8)
&       & 8.220(8)
&       & 8.243(8)
&8.274(9)\\
&6.586&" 
&       & 8.252(10)
&$\star$& 8.216(10)
&$\star$& 8.242(10)
&8.276(10)\\
&6.572&"
&       & 8.321(11)
&       & 8.288(10)
&       & 8.312(11)
&8.341(11)\\
&6.566&"
&       & 8.369(9)
&       & 8.335(9)
&       & 8.359(9)
&8.391(10)\\
\hline
$\approx 0.09$&7.11&0.0923
&       & 11.046(9)
&$\star$& 10.981(10)
&$\star$& 11.022(11)
&11.077(8)\\
&7.09&"
&$\star$& 11.020(10)
&       & 10.973(10)
&       & 11.006(12)
&11.040(10)\\
&7.08&"
&$\star$& 11.014(10)
&$\star$& 10.964(11)
&$\star$& 11.008(10)
&11.045(9)\\
\hline\hline
\end{tabular*}
\end{table*}

\subsection{Relativistic \emph{vs.}\ nonrelativistic operators}
\label{sec:relnonrel}

\begin{figure}[bp]
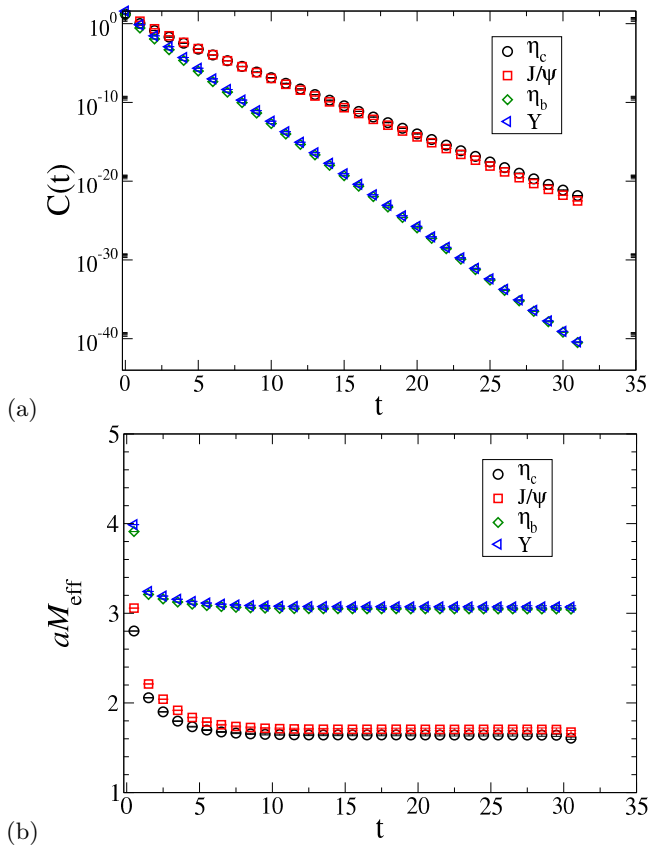

\centering
    (a)~\includegraphics*[width=8cm]{typical_props.eps}\\
    (b)~\includegraphics*[width=8cm]{effmass.eps}
    \caption{Propagators~(a) and effective masses~(b) 
        for the $\eta_c$, $J/\psi$, $\eta_b$, and $\Upsilon$ states, 
        with delta-function sources and sinks, 
        from the coarse ensemble with $am_l/am_s=0.01/0.05$.} 
\label{fig:prop}
\end{figure}

\begin{figure}[bp]
\centering
    \includegraphics*[width=8cm]{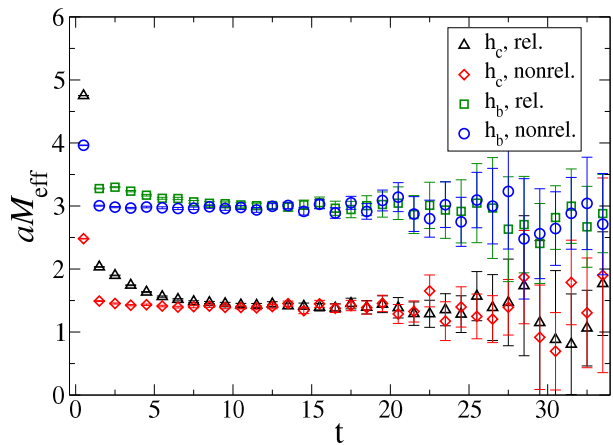}
    \caption{Comparison between effective masses for $h_c$ and $h_b$ 
        calculated with relativistic and nonrelativistic operators, 
        on the fine ensemble with $am_l/am_s=0.0124/0.031$.}
    \label{fig:1peff}
\end{figure}

The statistical quality of our data can be judged from 
Fig.~\ref{fig:prop}, which shows examples of typical two-point functions 
for the $1S$ pseudoscalar and vector states and their corresponding 
effective masses, calculated with relativistic operators.
The data are from the coarse ensemble with $am_l/am_s=0.01/0.05$. 
We have a clear signal and the effective masses have well-established 
plateaus.
As already mentioned, for the $1P$ states we used both relativistic and 
nonrelativistic types of operators.
Figure~\ref{fig:1peff} compares the effective masses of the $h_c(1P)$ 
and $h_b(1P)$ states, calculated with both types of operators.
The data show that the effective masses obtained with nonrelativistic 
operators plateau at an earlier $t_{\rm min}$.
Despite the fact that the statistics in the nonrelativistic case are, in 
this example, three times lower than in the relativistic case, the errors
on the fitted $h_c$ and $h_b$ masses are smaller than the ones 
calculated with relativistic operators.
This finding holds for all of the $1P$ states studied here.
Our statistics with the nonrelativistic operators are 2--3 times lower 
than with the relativistic ones (except for the medium-coarse case where 
they are 6 times lower), yet the errors on the masses are up to 50\% 
smaller, and in some cases smaller still---see
Tables~\ref{tab:cc-rel}--\ref{tab:bb-nonrel} for numerical comparisons.
The nonrelativistic operators couple much more weakly to the excited states
and, thus, yield effective mass plateaus of better quality and fitted 
masses with smaller errors.
All of our results for quarkonium masses are listed in 
Tables~\ref{tab:cc-rel}--\ref{tab:bb-nonrel} with statistical errors
calculated with the bootstrap method and symmetrized.

The central values of the $1P$ states calculated with relativistic and 
norelativistic operators occasionally differ by more than 1.5 uncorrelated 
$\sigma$.
This difference arises more often than expected, especially once 
correlations are considered.
In the tables, these cases are labeled with a star. 
To check whether this difference is due to statistics,
in some cases we carried out simultaneous fits to both the relativistic and
nonrelativistic correlators.
The masses extracted this way turned out to be indistinguishable from 
the masses from nonrelativistic data alone.
This was not surprising, because the data from relativistic sources had 
larger fluctuations than that from nonrelativistic sources.
Thus, in our further analysis of the chiral extrapolation and $a$ 
dependence, we use the nonrelativistic results for the $1P$ states 
wherever they are available.

\subsection{$\kappa$ tuning in quarkonium and heavy-strange mesons}
\label{sec:Qvsq}

\begin{figure}[b]
    \centering
    \includegraphics[width=0.45\textwidth]{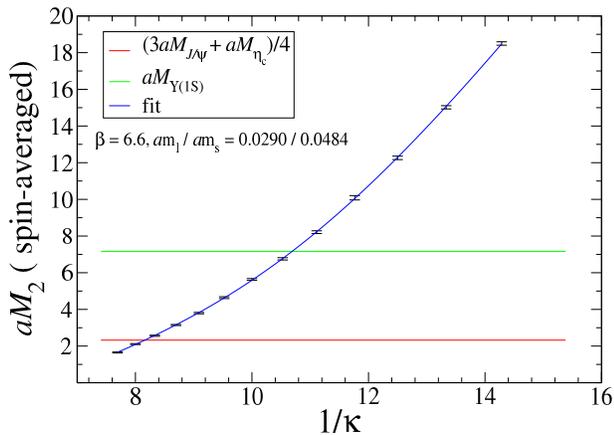}
    \caption[fig:tuning]{Spin-averaged kinetic mass $aM_2$ as a function of 
        $\kappa$, over a wide range, on the medium-coarse ensemble
        with $am_l/am_s=0.0290/0.0484$.
        With a polynomial fit to the data, we can read off
        $\quarter(aM_{\eta_c}+3aM_{J/\psi})$ and $aM_{\Upsilon}$, finding
        $\kappa_c\approx0.122$ and $\kappa_b\approx0.094$.}
    \label{fig:tuning} 
\end{figure}

In Sec.~\ref{sec:heavy}, we argued that the best way to tune the 
hopping parameter~$\kappa$ is to use the spin-averaged kinetic mass of 
heavy-strange hadrons.
If instead one would tune to the kinetic mass of the (spin-averaged) 
quarkonium ground state, the resulting tuned $\kappa$ could be 
different at nonzero lattice spacing.
To study this discrepancy we have computed the quarkonium 
$\overline{1S}$ kinetic mass for a wide range of~$\kappa$ on the 
medium-coarse ensemble with $am_l/am_s=0.0290/0.0484$.
Figure~\ref{fig:tuning} shows the results and also shows the 
physical $aM(\overline{1S})$ and $aM_{\Upsilon}$.%
\footnote{When this tuning was carried out, the $\eta_b$ had not yet
been observed by experiment.}
From a polynomial fit to the data, we get $\kappa_c\approx0.122$ for
the charmed quark, which is the same value as the one we obtain from
matching to $D_s$.
However, because the relevant discretization effects are larger
in bottomonium than in $B$~mesons, the tuned values of the hopping
parameter differ substantially: $\kappa_b\approx0.094$ from $\Upsilon$
\emph{vs.}\ $\kappa_b\approx0.076$ from~$B_s$.

When we tune to the $D_s$, some uncertainty in $\kappa$ arises.
We take the tuning error in $\kappa_c$ to be 0.0015 and 
in $\kappa_b$ to be~0.006.
Reference~\cite{Freeland:2009} finds uncertainties (statistical and 
fitting) in this range on the medium-coarse, coarse, and fine ensembles,
and here we assume the same for the extra-coarse ensembles.
We discuss in the next subsection how to propagate these errors to our 
computed splittings.

Above we mentioned a small difference in tuning the clover coupling 
for the coarse ensembles.
The value of the tadpole coefficient $u_0$ used in that analysis was 
determined from mean Landau gauge link whereas the coefficient used in 
the others was determined from the plaquette.
This difference means that our bare quark mass, \emph{i.e.}, $\kappa$, 
has a slightly different definition on the coarse ensembles.
Discrepancies in mass splittings caused by this choice should be 
eliminated via the nonperturbative tuning.

\subsection{$\kappa$-tuning uncertainties}
\label{sec:kappa}

Tables~\ref{tab:cc-rel}--\ref{tab:bb-nonrel} and most of the plots in 
Sec.~\ref{sec:results} show statistical errors only, because the 
foremost aim of this paper is to understand the pattern of 
discretization errors.
A systematic error also arises from inaccuracies in tuning~$\kappa_c$ 
and $\kappa_b$, and to study the continuum limit it is necessary 
to propagate this error to the mass splittings.
We discuss here how we treat these uncertainties.

Several pieces of evidence show that the spin-averaged splittings 
depend very little on $\kappa$.
These splittings vary little from charmonium to 
bottomonium~\cite{Amsler:2008zz}, a feature understood to be a 
consequence of both systems lying between the confining and Coulombic 
part of the potential~\cite{Quigg:1979vr,Kwong:1987mj}.
This feature is, in fact, reproduced in our lattice-QCD data.
Moreover, earlier work in the quenched approximation~%
\cite{ElKhadra:1992ir} and with $n_f=2$~\cite{ElKhadra:2000zs}  
show negligible $\kappa$ dependence for spin-averaged splittings.
Thus, we shall assume that the $\kappa$-tuning error for these 
splittings can be neglected.

For spin-dependent splittings, we compute the $1S$ hyperfine
splitting as a function of $\kappa$, on the medium-coarse 
ensemble with $am_l/am_s=0.0290/0.0484$, the same ensemble as in 
Fig.~\ref{fig:tuning}.
The data are summarized in Table~\ref{tab:HFSvskappa}.
\begin{table}[tp]
    \caption{$1^3\!S_1$-$1^1\!S_0$ hyperfine splittings in $r_1$ 
        units as a function of the valence $\kappa$ calculated for the 
        medium-coarse ensemble with $am_l/am_s=0.0290/0.0484$.}
    \label{tab:HFSvskappa} 
    \begin{tabular}{cd}
        \hline\hline
            $\kappa$   & 
            \multicolumn{1}{c}{$\hspace*{3em}r_1[M(1^3\!S_1)-M(1^1\!S_0)]$} \\
        \hline
            0.070 & 0.0247(9)  \\
            0.075 & 0.0299(11) \\
            0.080 & 0.0369(12) \\
            0.085 & 0.0442(13) \\
            0.090 & 0.0531(14) \\
            0.095 & 0.0631(15) \\
            0.100 & 0.0749(17) \\
            0.105 & 0.0885(18) \\
            0.110 & 0.1029(23) \\
            0.115 & 0.1244(27) \\
            0.120 & 0.1499(33) \\
            0.125 & 0.1836(40) \\
            0.130 & 0.2335(49) \\
        \hline\hline
    \end{tabular}
\end{table}
\begin{figure*}
\centering
    (a) \includegraphics*[width=8cm]{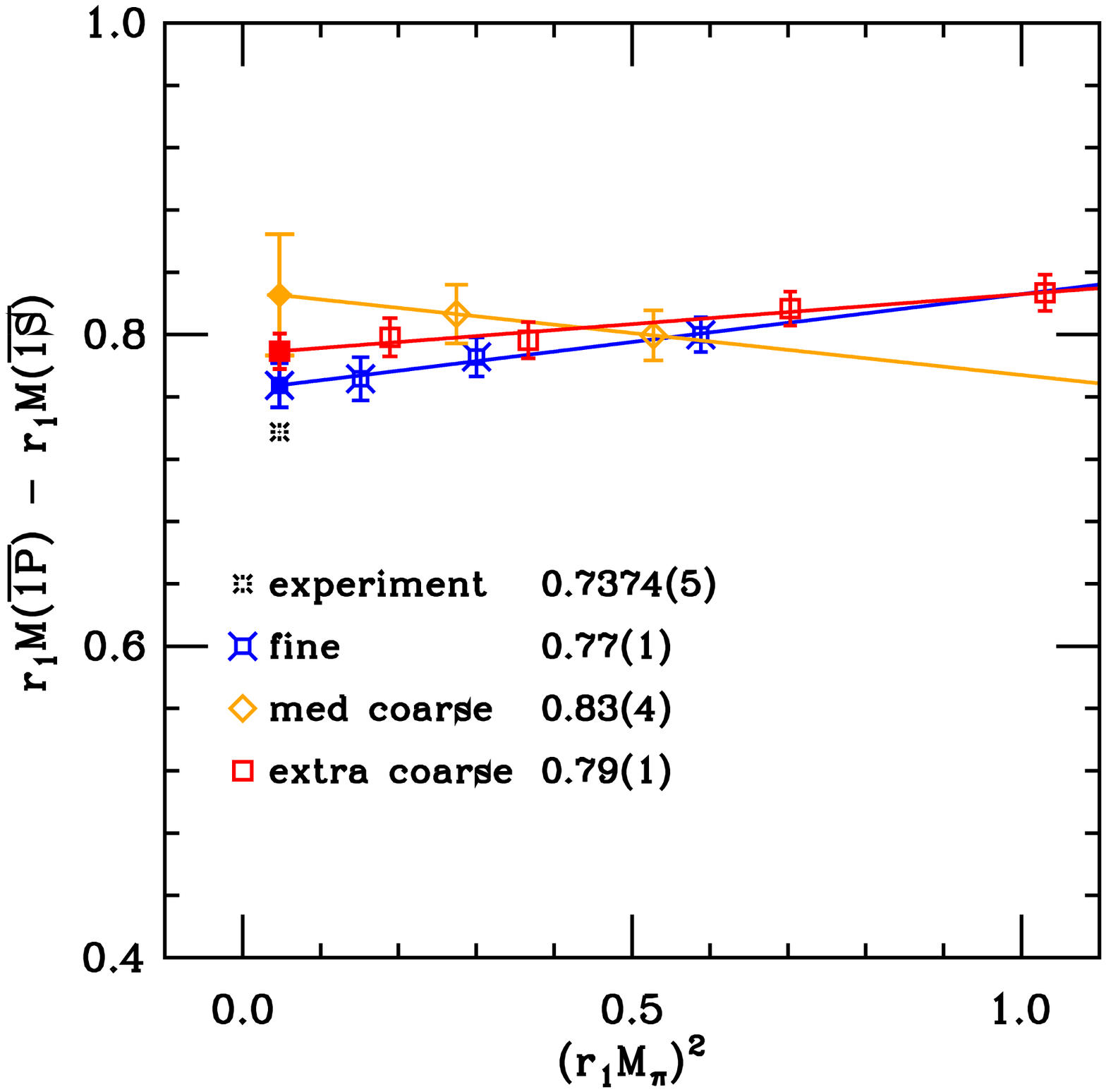}\hfill
    (b) \includegraphics*[width=8cm]{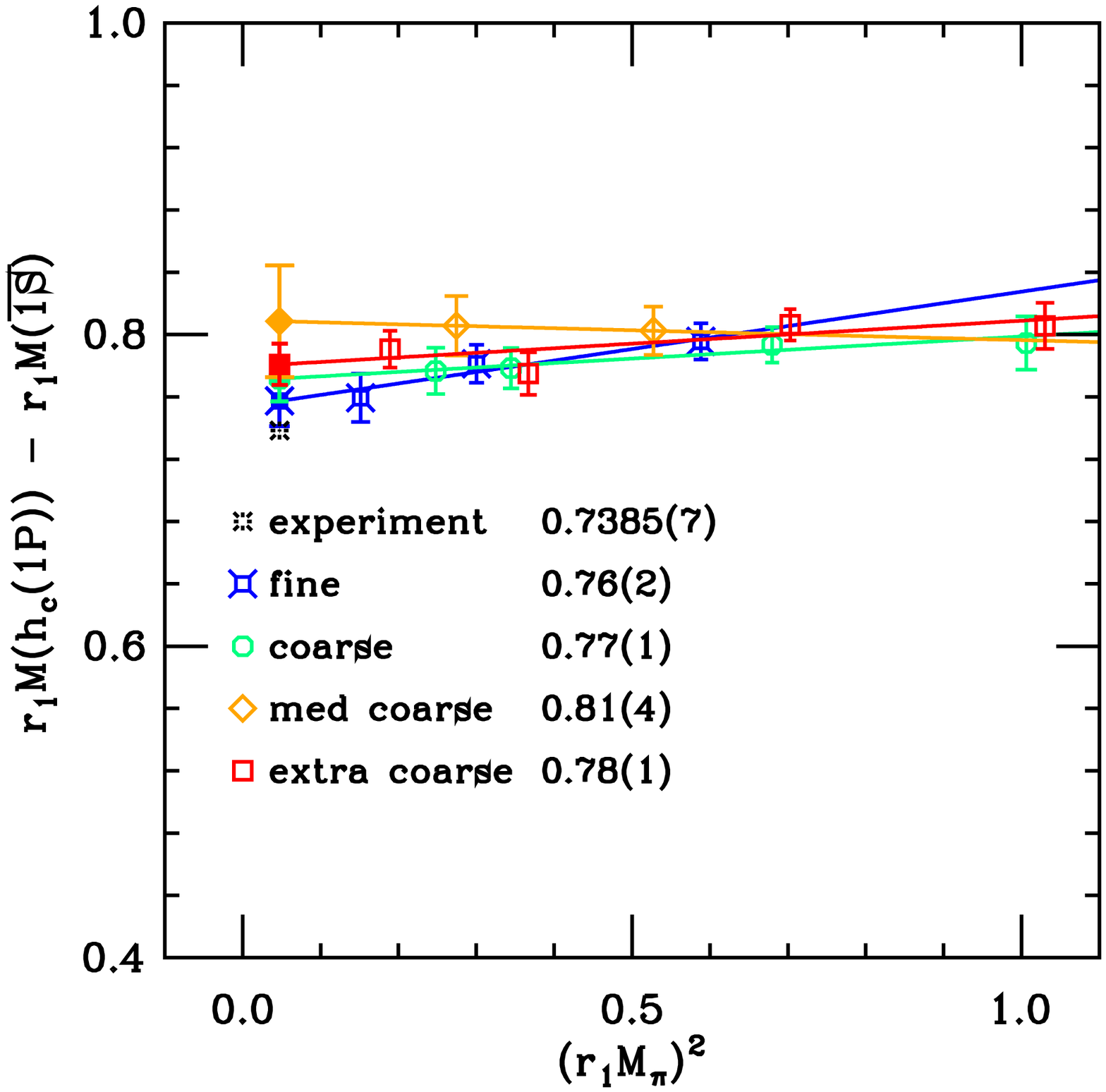}
    \caption{The (a) $\overline{1P}$\/-$\overline{1S}$ and
        (b) $1^1\!P_1$-$\overline{1S}$ splittings in charmonium.
        The fine ensemble data are in blue fancy squares, the coarse
        in green circles, the medium-coarse in orange diamonds and the 
        extra-coarse in red squares.
        The chirally extrapolated values are given in the legend and 
        plotted with filled symbols.}
    \label{fig:cc1p-1s}
\end{figure*}
We fit the data to the form
\begin{eqnarray}
    \mu & = & 1/\kappa - 1/\kappa_{\rm cr} , \\
    \hspace*{-2em}
    \mathrm{HFS} & = & b_0/\mu^2 + b_1/\mu^3 + b_2/\mu^4 +
        b_3/\mu^5 + b_4/\mu^6 ,\hspace*{1em}
\end{eqnarray}
for $\kappa_{\rm cr} = 0.145$, which enforces the requirement that, at
large heavy quark mass $m_0=\mu/2a$, the splitting goes as $1/m_0^2$.
The fit gives $\chi^2/\mathrm{dof} = 0.6/8$.
From the fit result we estimate that an error of 0.0015 in the 
determination of $\kappa_c$ results in a 6\% error in the charmonium 
hyperfine splitting, and an error of 0.006 in the determination of 
$\kappa_b$, a 22\% error in the bottomonium hyperfine splitting.
We expect that these errors are characteristic of all splittings driven 
by the spin-spin and tensor terms in the quarkonium effective potential, 
since in the nonrelativistic treatment, they all stem from the same 
term in the heavy-quark effective Lagrangian.

The spin-orbit splitting remains to be considered.
In our data and in experiment, it decreases from charmonium to 
bottomium similarly to the hyperfine and tensor splittings.
Therefore, we shall assume the same relative error from the uncertainty 
in tuning $\kappa$.

Below we also present results for the splittings between twice the
spin-averaged mass of $D_s$ and $D_s^*$, and of $B_s$ and $B_s^*$,
and the corresponding $\overline{1S}$ quarkonium mass.
To estimate their $\kappa$-tuning errors we have calculated these 
spin-averaged masses for several values of $\kappa$ near $\kappa_c$ and 
$\kappa_b$ on the coarse ensemble with $am_l/am_s=0.01/0.05$.
These direct measurements allow us to propagate the $\kappa$-tuning errors 
from the masses to the mass splittings.
We obtain an error of 1.3\% for charm and 13\% for bottom.
We assume the same error for these splittings at other lattice 
spacings.

\section{Spectrum Results}
\label{sec:results}

We now present plots of quarkonium mass splittings as a function of the 
square of the sea-quark pion mass.
The splittings and their errors are calculated using the bootstrap method.
In most cases, we expect the dependence on the sea-quark mass to be mild, 
so we perform on our results a chiral extrapolation linear in $M_\pi^2$ 
down to the physical pion.
The extrapolated values are denoted in each plot with filled symbols.
The error bars come from symmetrizing the 1$\sigma$ (68\%) interval of 
the bootstrap distribution.

Where possible, we compare our results to experimental measurements.
As a rule we take the average values from the compilation of the 
Particle Data Group~\cite{Amsler:2008zz}.
The exception is the mass of the $\eta_b(1S)$ meson, which has only 
recently been observed.
We take $M_{\eta_b}=9390.9\pm2.8$~MeV, based on our average of two 
measurements by the BaBar Collaboration~\cite{:2008vj,Aubert:2009pz} 
and one by the CLEO Collaboration~\cite{Bonvicini:2009hs}.

In examining the results, we are interested in seeing how well 
we can understand discretization errors via the nonrelativistic 
description of Eqs.~(\ref{eq:SequivLeff})--(\ref{eq:Leff4}).
We therefore carry out separate chiral extrapolations at each lattice 
spacing, and discuss whether the $a$ dependence, and any deviations from 
experiment, make sense.

From the effective Lagrangian discussion, we expect different 
discretization errors to affect spin-averaged and spin-dependent 
splittings.
Errors in the spin-averaged splittings stem from the Darwin 
($\bm{D}\cdot\bm{E}$) term and the two $p^4$ terms.
Errors in the spin-dependent splittings stem from the chromomagnetic
($i\bm{\sigma}\cdot\bm{B}$) and spin-orbit 
($i\bm{\sigma}\cdot\bm{D}\times\bm{E}$) terms.
Moreover, from the general structure of potentials arising from 
QCD~\cite{Eichten:1980mw,Peskin:1983up}, we learn that 
$i\bm{\sigma}\cdot\bm{B}$ predominantly affects
$M(nS_{\textrm{HFS}})$ and $M(nP_{\textrm{tensor}})$, while
$i\bm{\sigma}\cdot\bm{D}\times\bm{E}$ affects 
$M(nP_{\textrm{spin-orbit}})$.

\begin{figure*}
\centering
    (a) \includegraphics*[width=8cm]{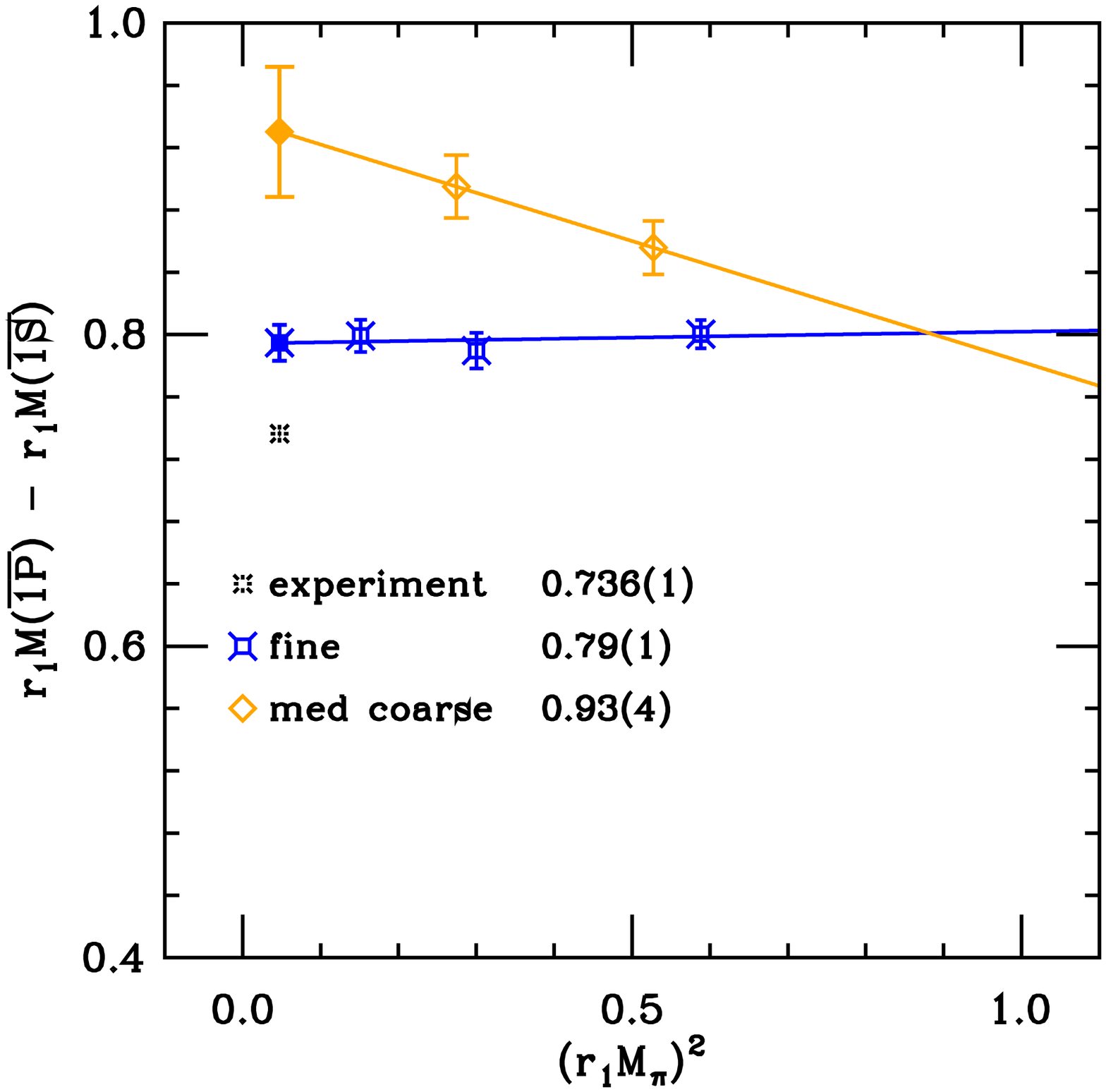}\hfill
    (b) \includegraphics*[width=8cm]{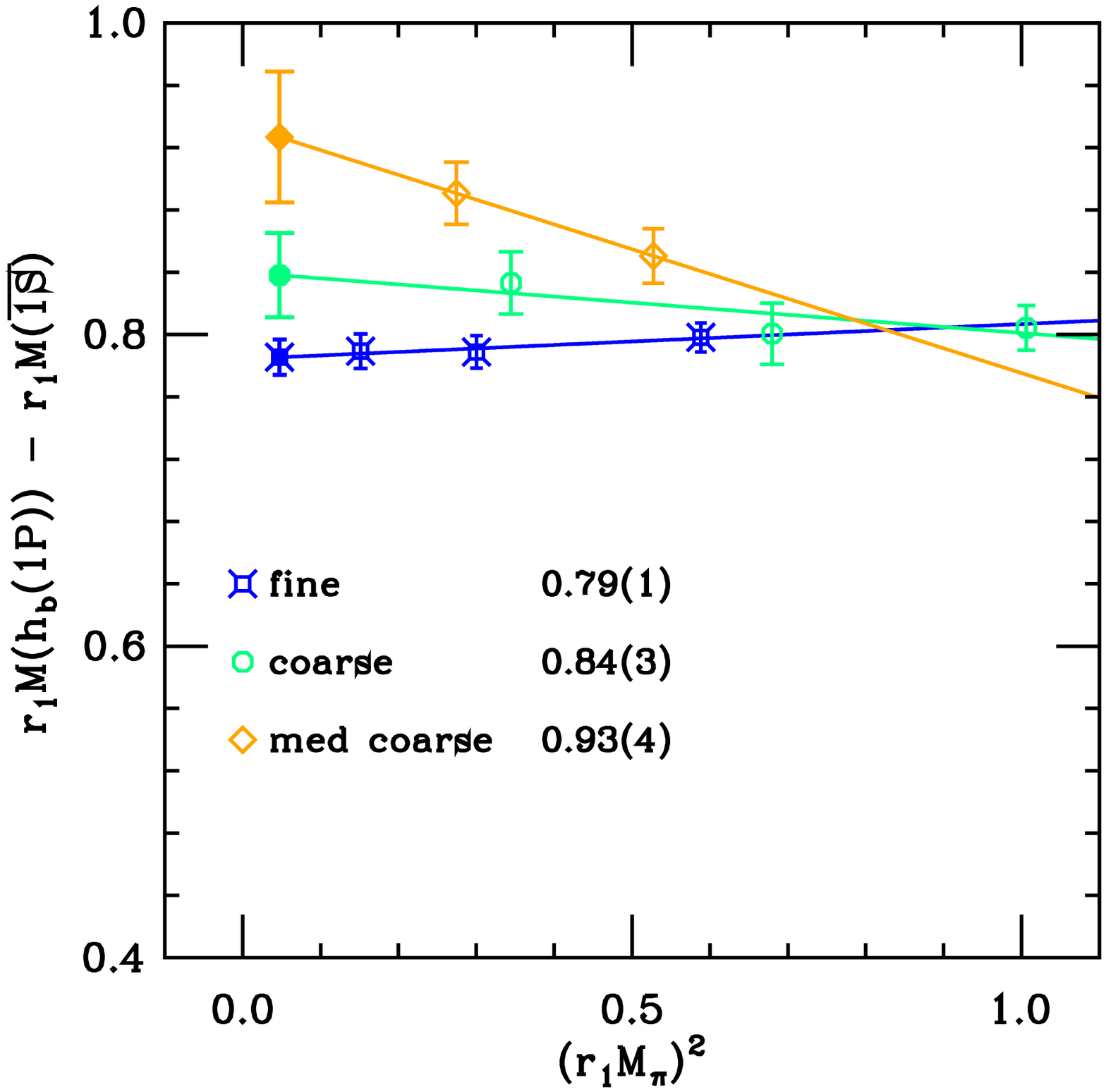}
    \caption{The (a) $\overline{1P}$\/-$\overline{1S}$ and (b)
        $1^1\!P_1$-$\overline{1S}$ splittings in bottomonium.
        Color code as in Fig.~\ref{fig:cc1p-1s}.}   
    \label{fig:bb1p-1s}
\end{figure*}
\begin{figure*}
\centering
    (a)~\includegraphics*[width=8cm]{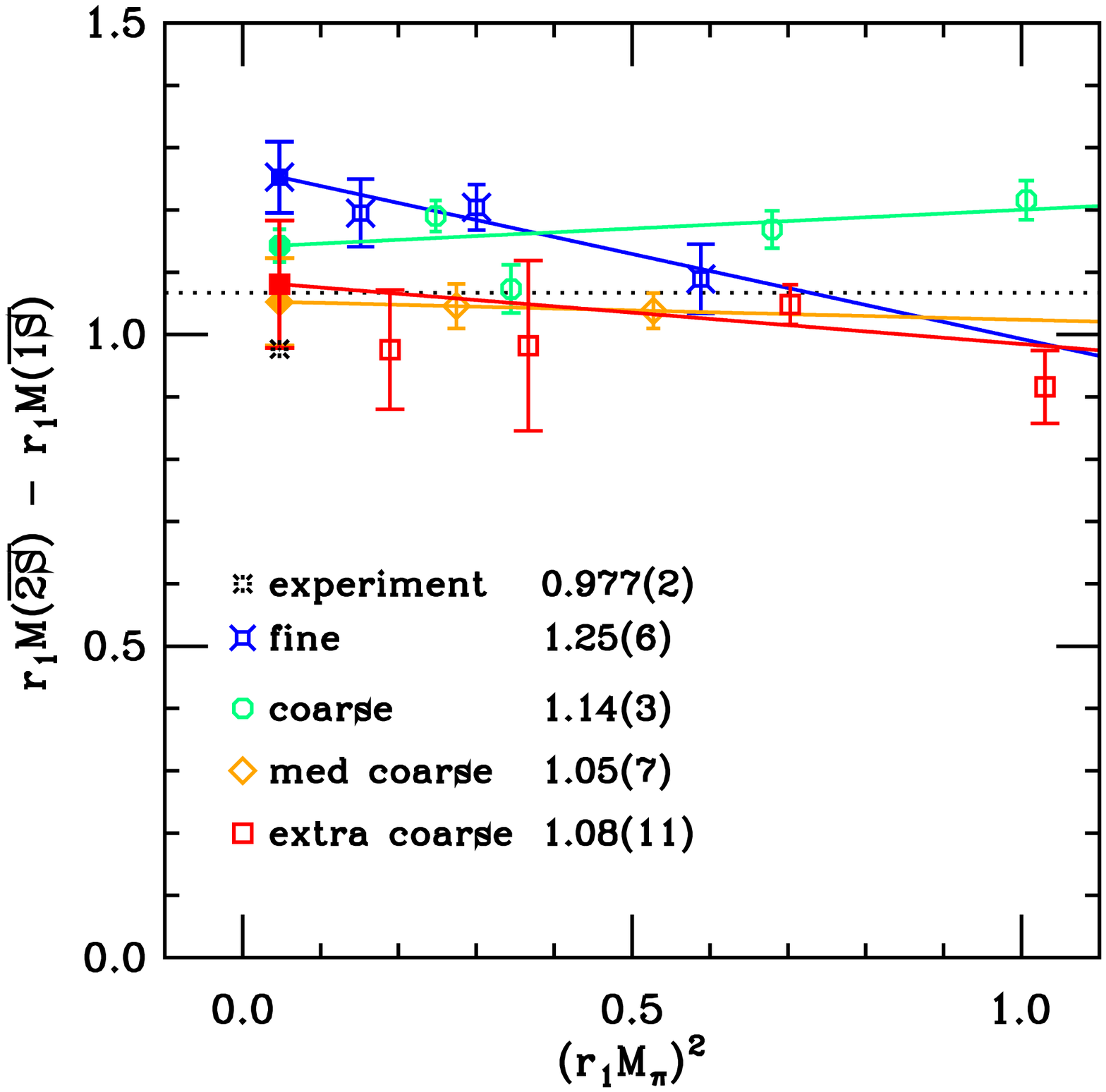}\hfill
    (b)~\includegraphics*[width=8cm]{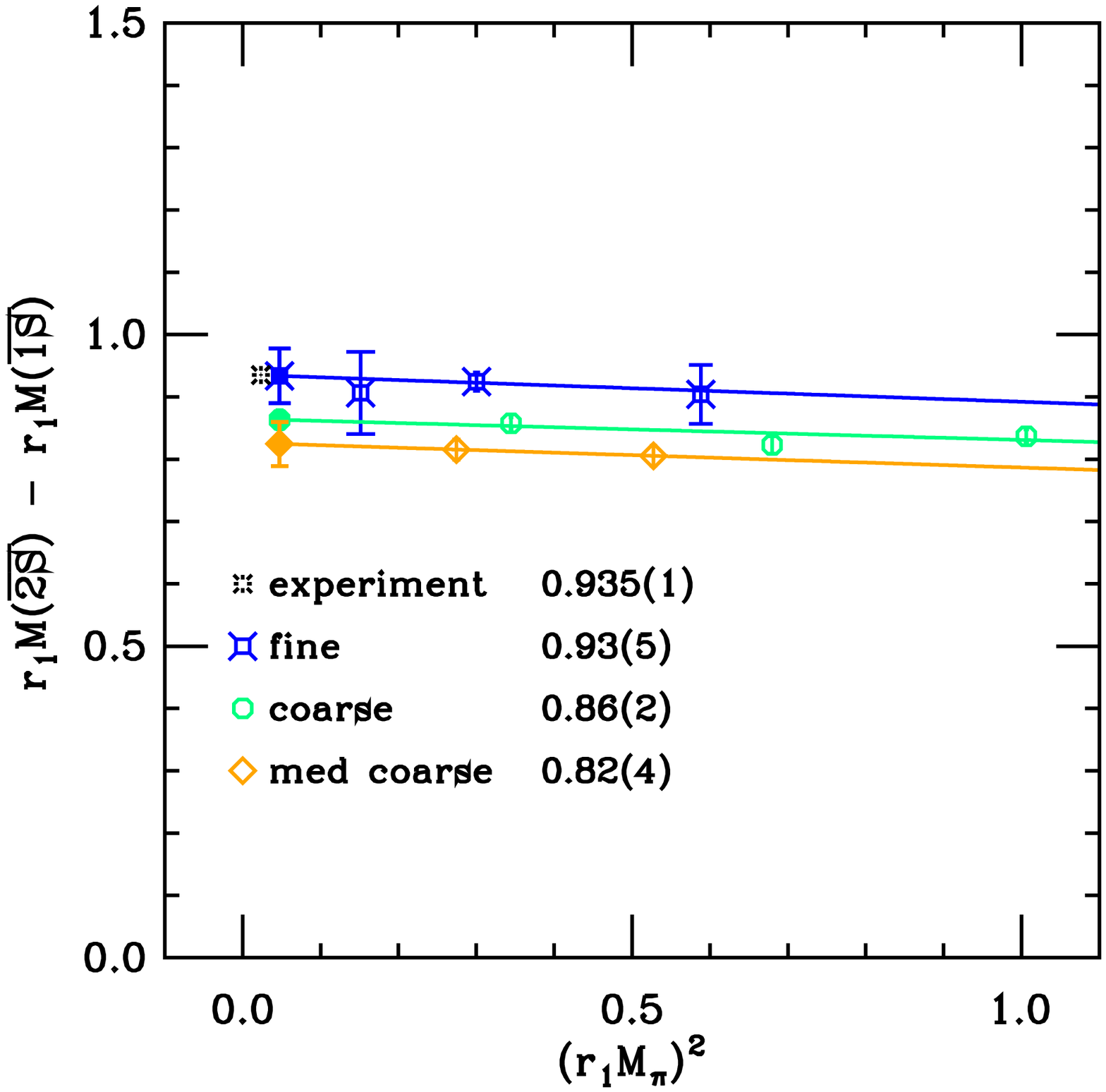}
    \caption{Splitting between the $\overline{2S}$ and $\overline{1S}$ 
        levels of (a)~charmonium, (b)~bottomonium.
        The dotted line in (a) indicates the open-charm threshold.
        The experimental point in (b) is \emph{not} the spin-averaged
        splitting, but the $\Upsilon(2S)$-$\overline{1S}$ mass 
        difference, since the $\eta^\prime_b$ has not been observed.}
    \label{fig:2s-1s}
\end{figure*}

\subsection{Spin-averaged splittings}


Let us start with $\overline{1P}$\/-$\overline{1S}$ and 
$1^1\!P_1$-$\overline{1S}$ splittings, plotted in Figs.~\ref{fig:cc1p-1s}
and~\ref{fig:bb1p-1s} \emph{vs}.~$(r_1M_\pi)^2$.
In the nonrelativistic picture, they arise predominantly at order~$v^2$ 
via the kinetic energy, which our tuning of $\kappa$ should normalize 
correctly.
The spin-dependent terms in $\mathcal{L}_{\rm HQ}^{(4)}$ 
[cf.\ Eq.~(\ref{eq:Leff4})] do not contribute to spin averages
($\overline{1S}$, $\overline{1P}$) or to a spin singlet ($1^1\!P_1$).
Discretization errors remain, however, at order~$v^4$ via the mismatches 
in Eqs.~(\ref{eq:mEm2})--(\ref{diff:w4}).
We assess these results using the error estimates in 
Ref.~\cite{Oktay:2008ex}, which account for both the $a$ dependence and 
the relative $v^2$ suppression.
\begin{figure*}
\centering
    \includegraphics*[width=8cm]{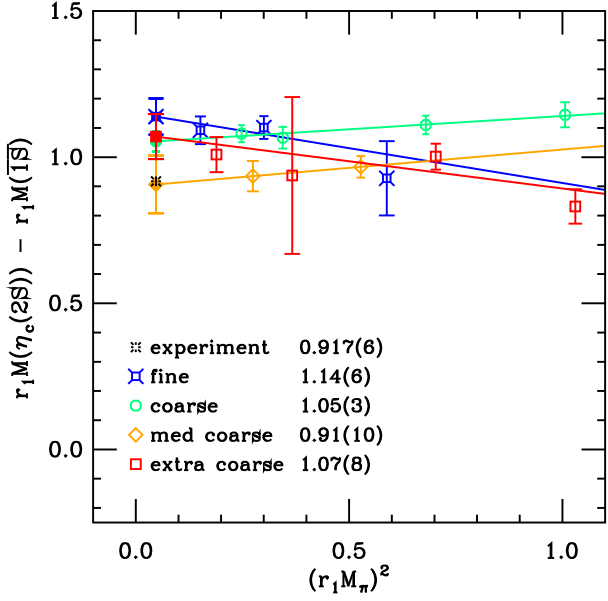}\hfill
    \includegraphics*[width=8cm]{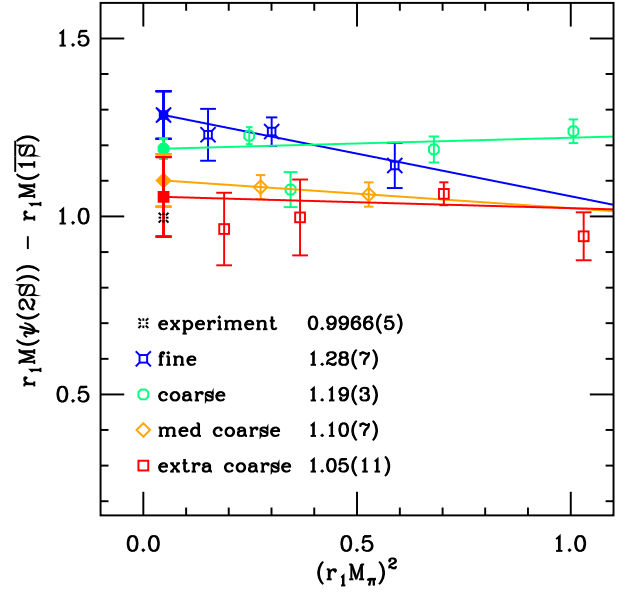}
    \caption{Splittings in charmonium between the individual $2S$ states 
        and the $\overline{1S}$ level.}
    \label{fig:cc2s-1s}
\end{figure*}
\begin{figure*}
\centering
    \includegraphics*[width=8cm]{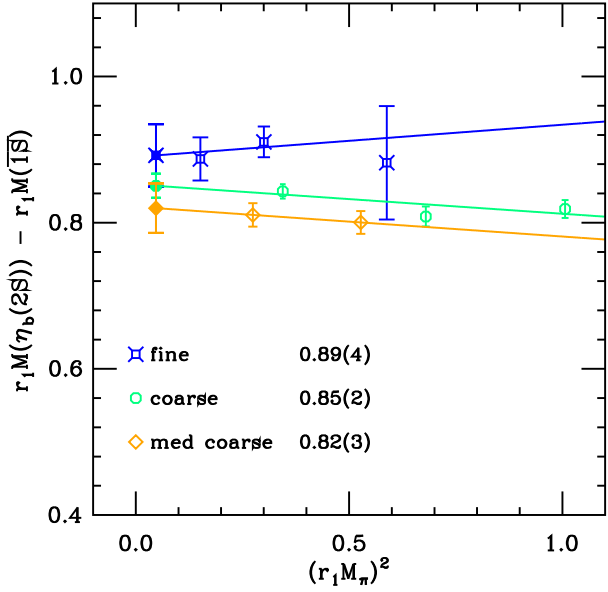}\hfill
    \includegraphics*[width=8cm]{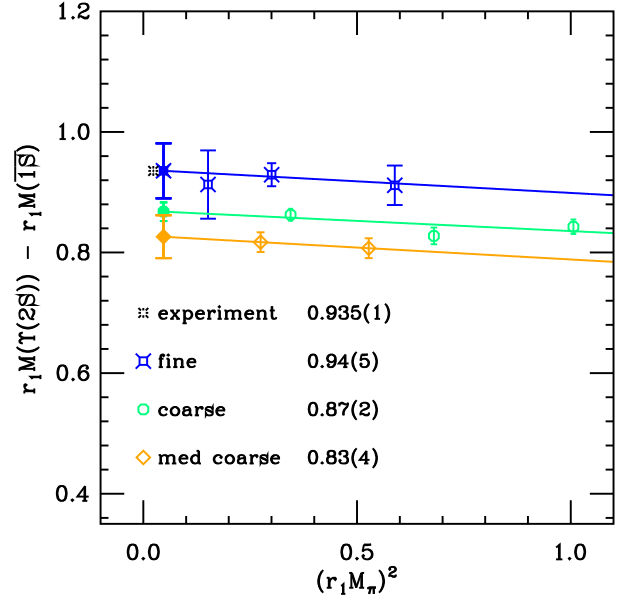}
    \caption{Splittings in bottomonium between the individual $2S$ states 
        and the $\overline{1S}$ level.}
    \label{fig:bb2s-1s}
\end{figure*}

Our results for charmonium are shown in Fig.~\ref{fig:cc1p-1s}.
Our results for both splittings approach the continuum physical point 
as the lattice spacing decreases, and the size of the discretization 
effects is about what one expects: 5--6\% from $m_E\neq m_2$ and 3--6\% 
from $m_4\neq m_2$ \cite{Oktay:2008ex}.

The $\overline{1P}$\/-$\overline{1S}$ and $1^1\!P_1$-$\overline{1S}$ 
splittings in bottomonium are given in Fig.~\ref{fig:bb1p-1s}.
These splittings agree acceptably with experiment, given the estimated 
discretization errors, 2--3\% from $m_E\neq m_2$ and 2--5\% from 
$m_4\neq m_2$ \cite{Oktay:2008ex}.
We cannot compare the $h_b(1^1\!P_1)$ mass with experiment, because 
that state has not been observed~\cite{Amsler:2008zz}, but our results 
for the $1^1\!P_1$ level agree very well with the $\overline{1^3\!P_J}$ 
average.
\begin{figure*}
\centering
    (a) \includegraphics*[width=8cm]{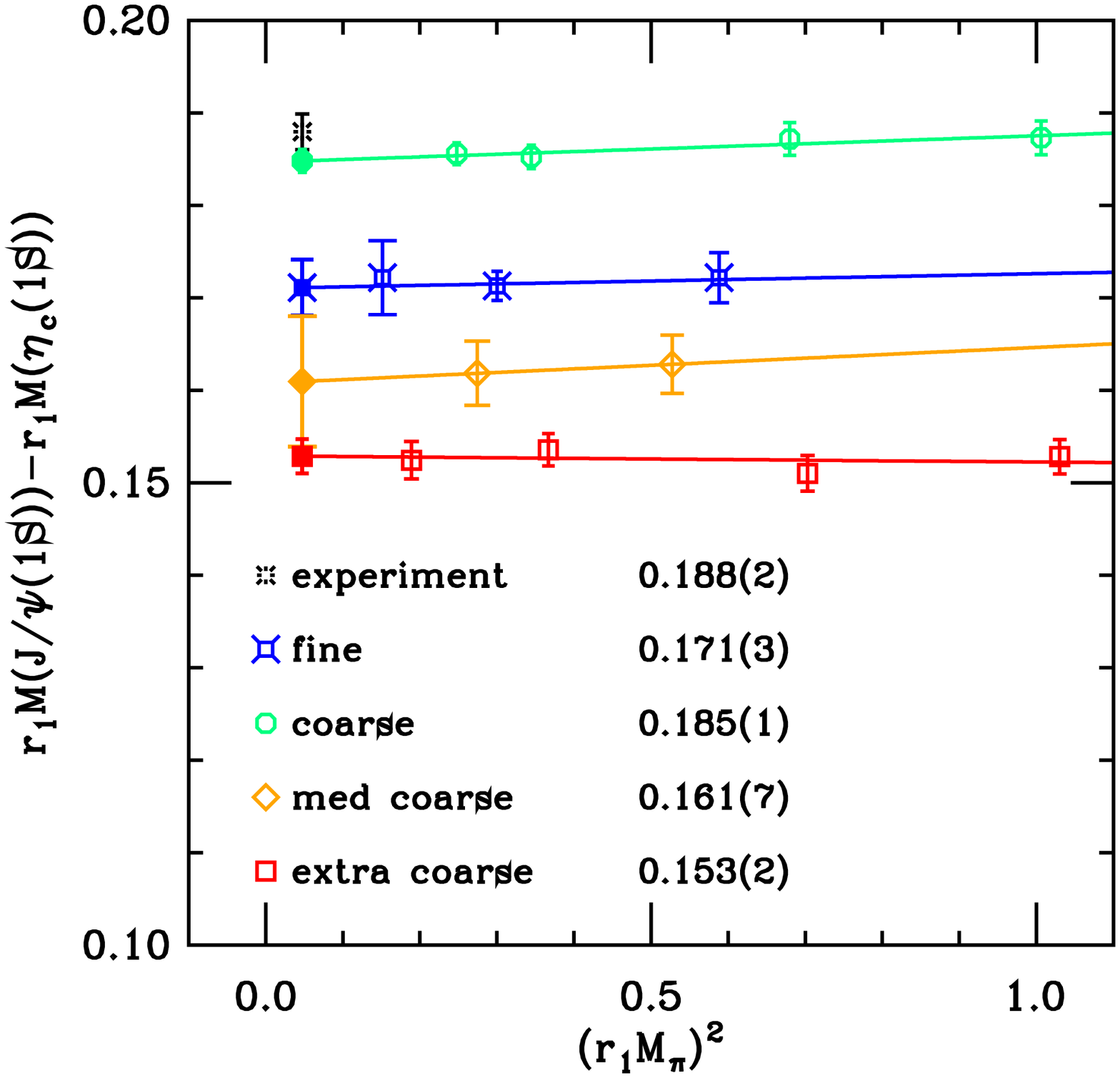}\hfill
    (b) \includegraphics*[width=8cm]{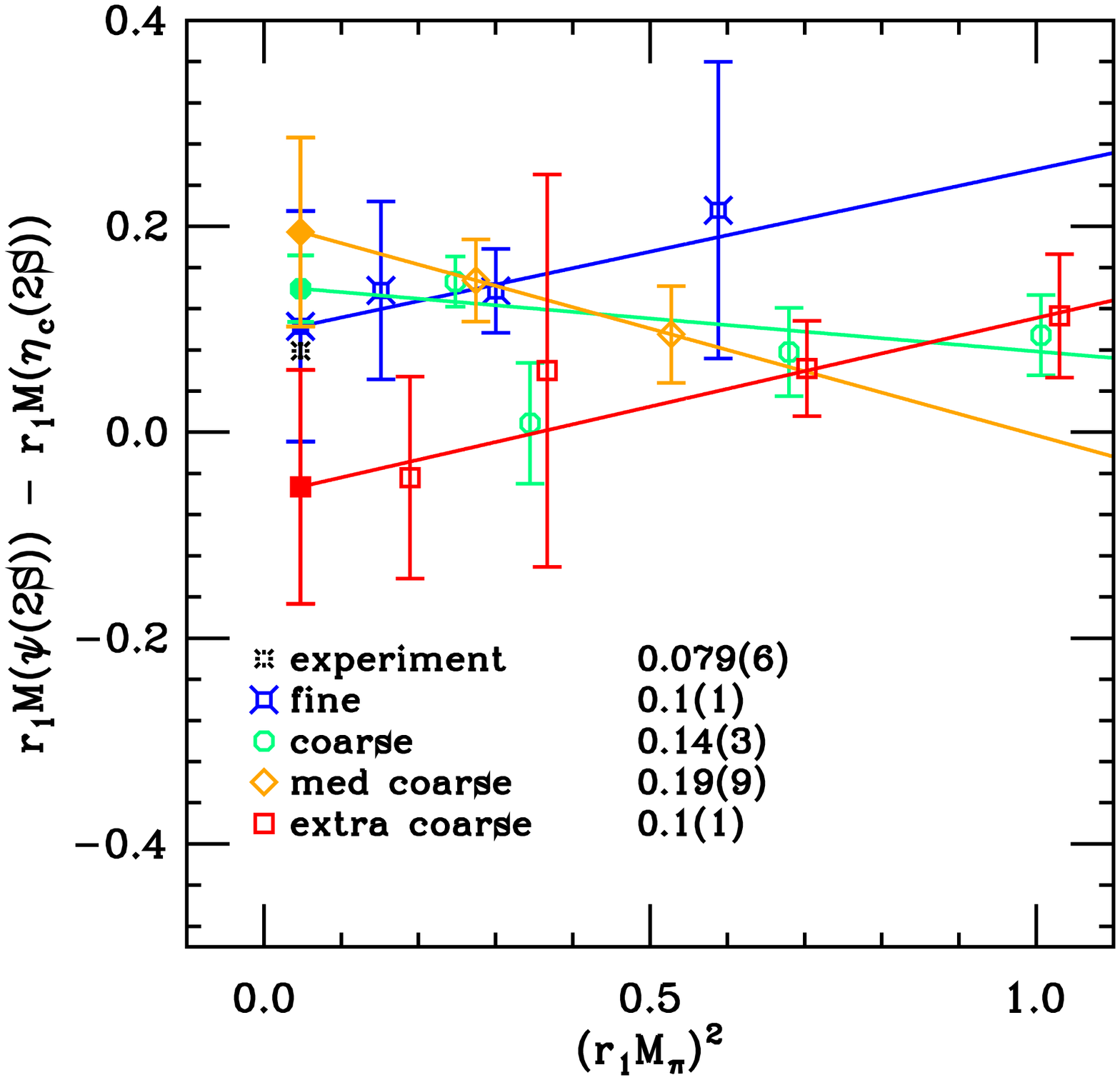}
    \caption{Charmonium hyperfine splittings for (a) $1S$, (b)~$2S$.}
    \label{fig:cc_hyp}
\end{figure*}
\begin{figure*}
\centering
    (a) \includegraphics*[width=8cm]{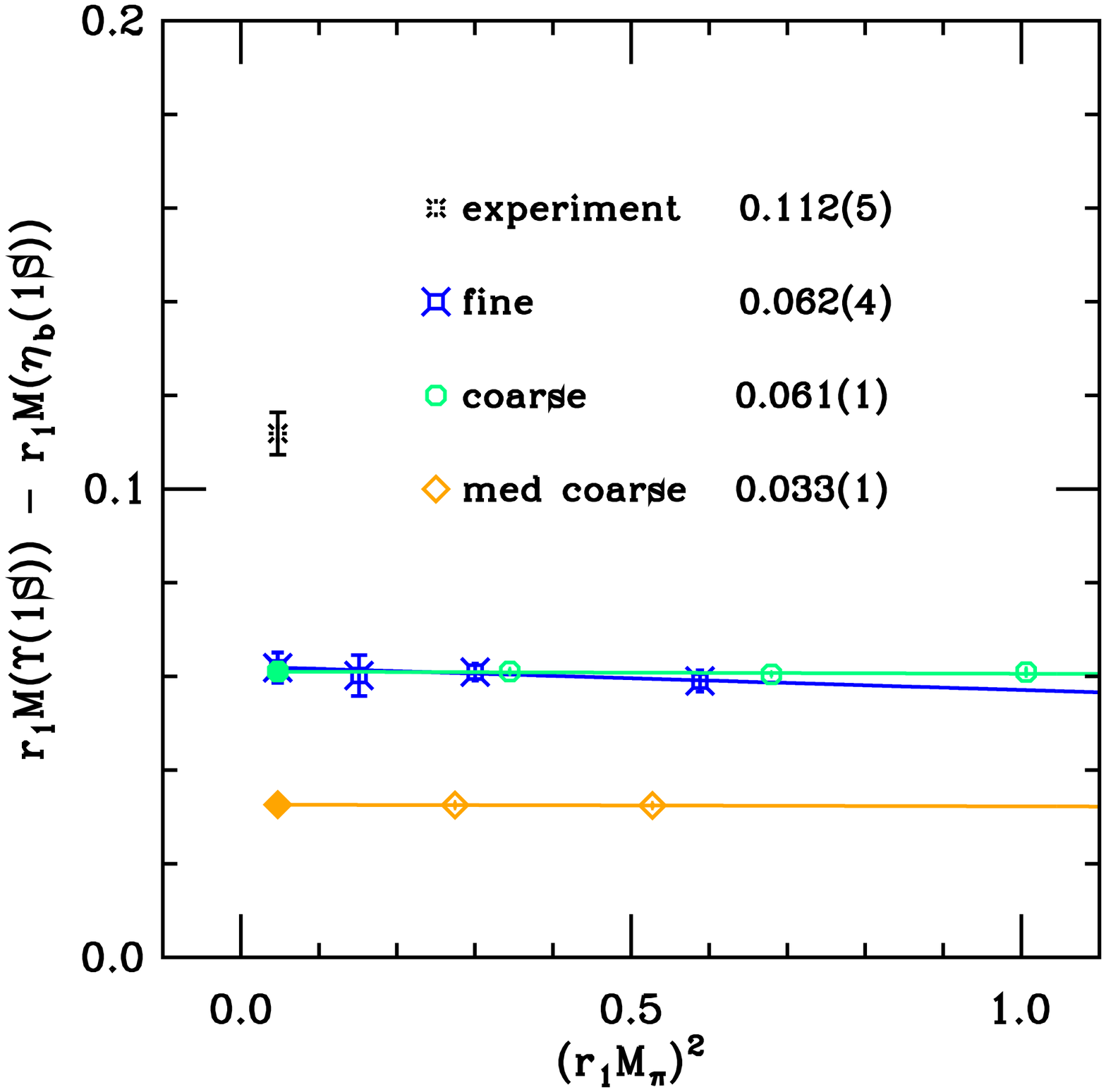}\hfill
    (b) \includegraphics*[width=8cm]{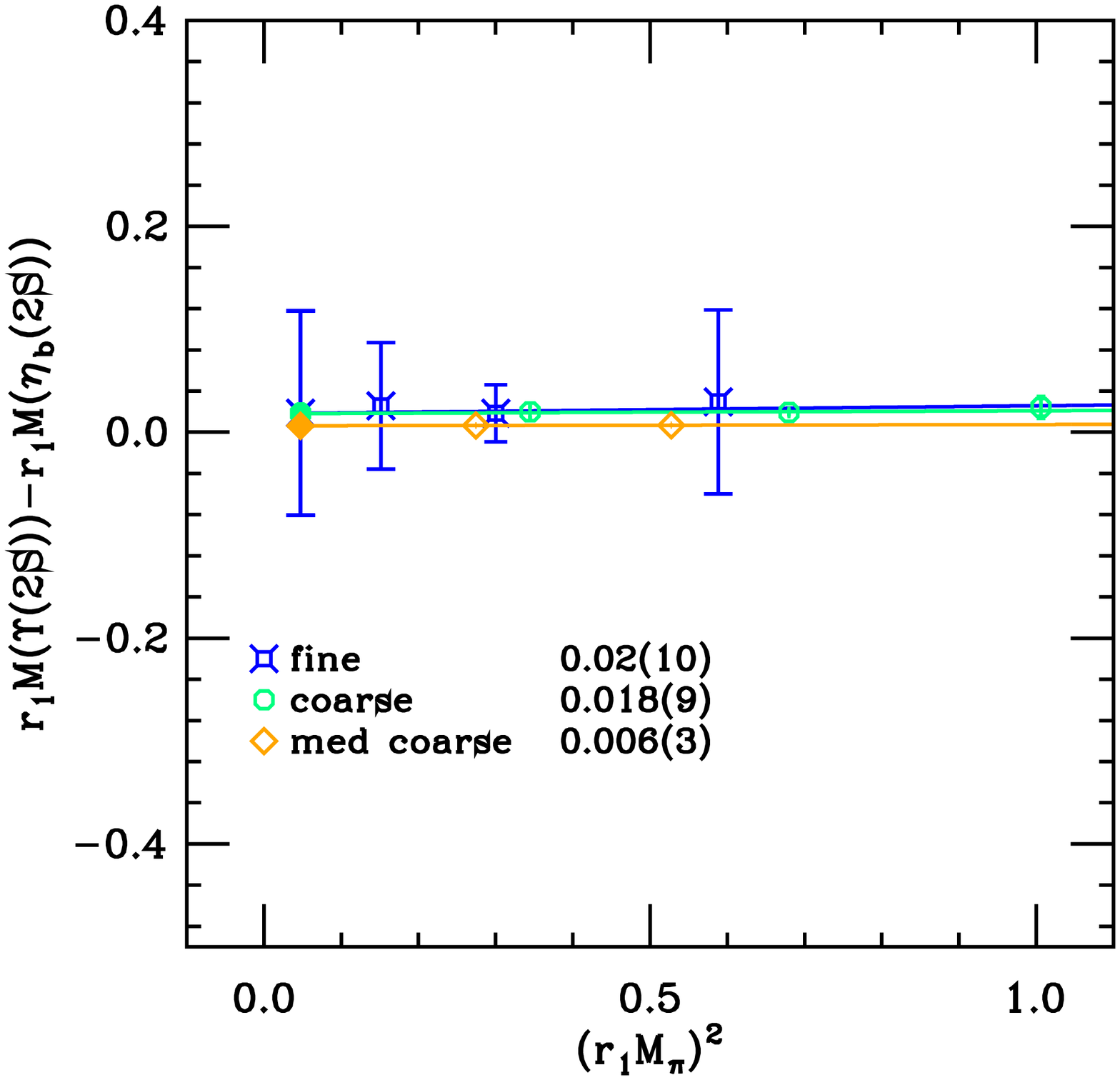}
    \caption{Bottomonium hyperfine splittings for (a) $1S$, (b)~$2S$.}
    \label{fig:bb_hyp}
\end{figure*}
\begin{figure*}
\centering
    (a) \includegraphics*[width=8cm]{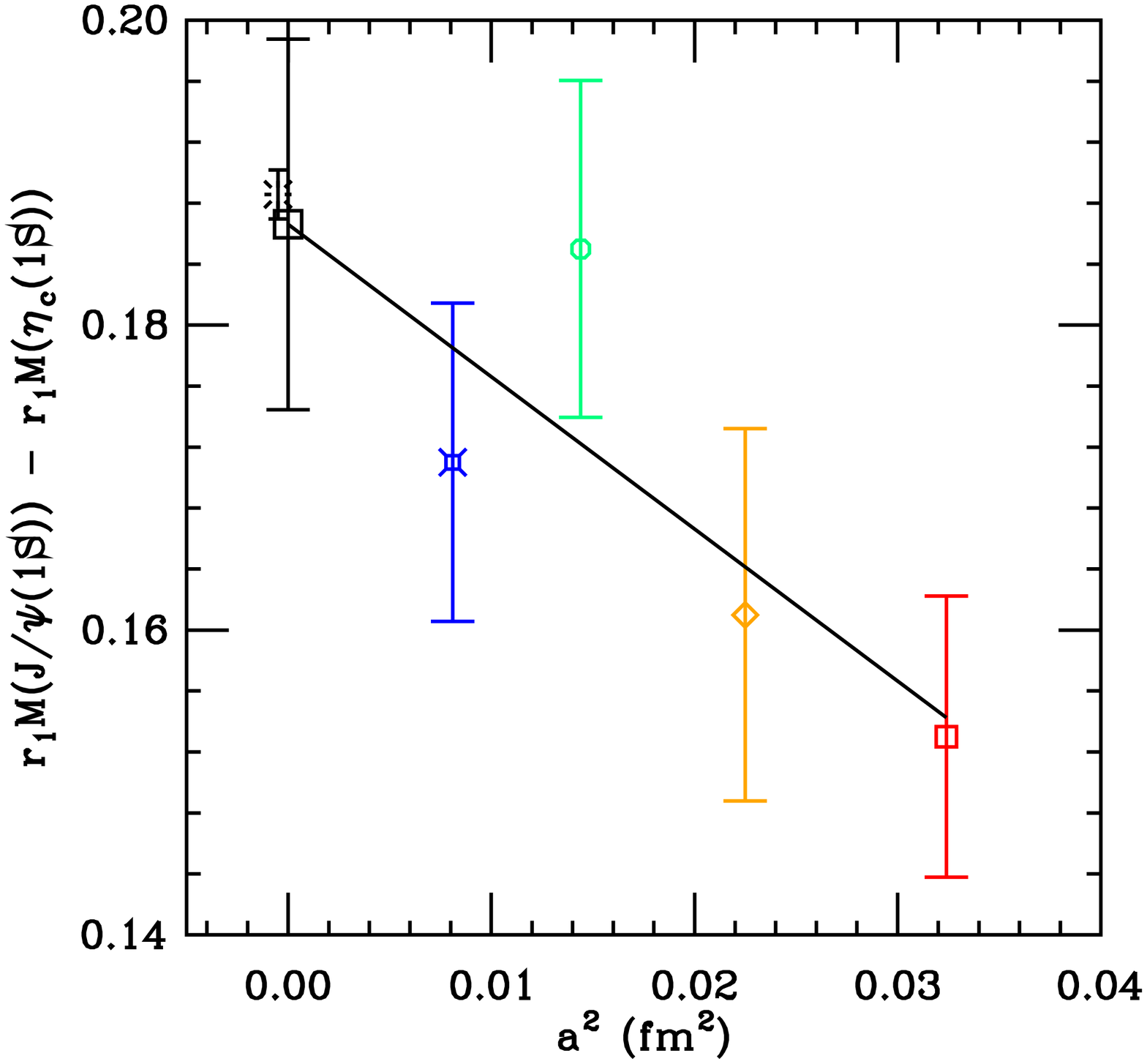}\hfill
    (b) \includegraphics*[width=8cm]{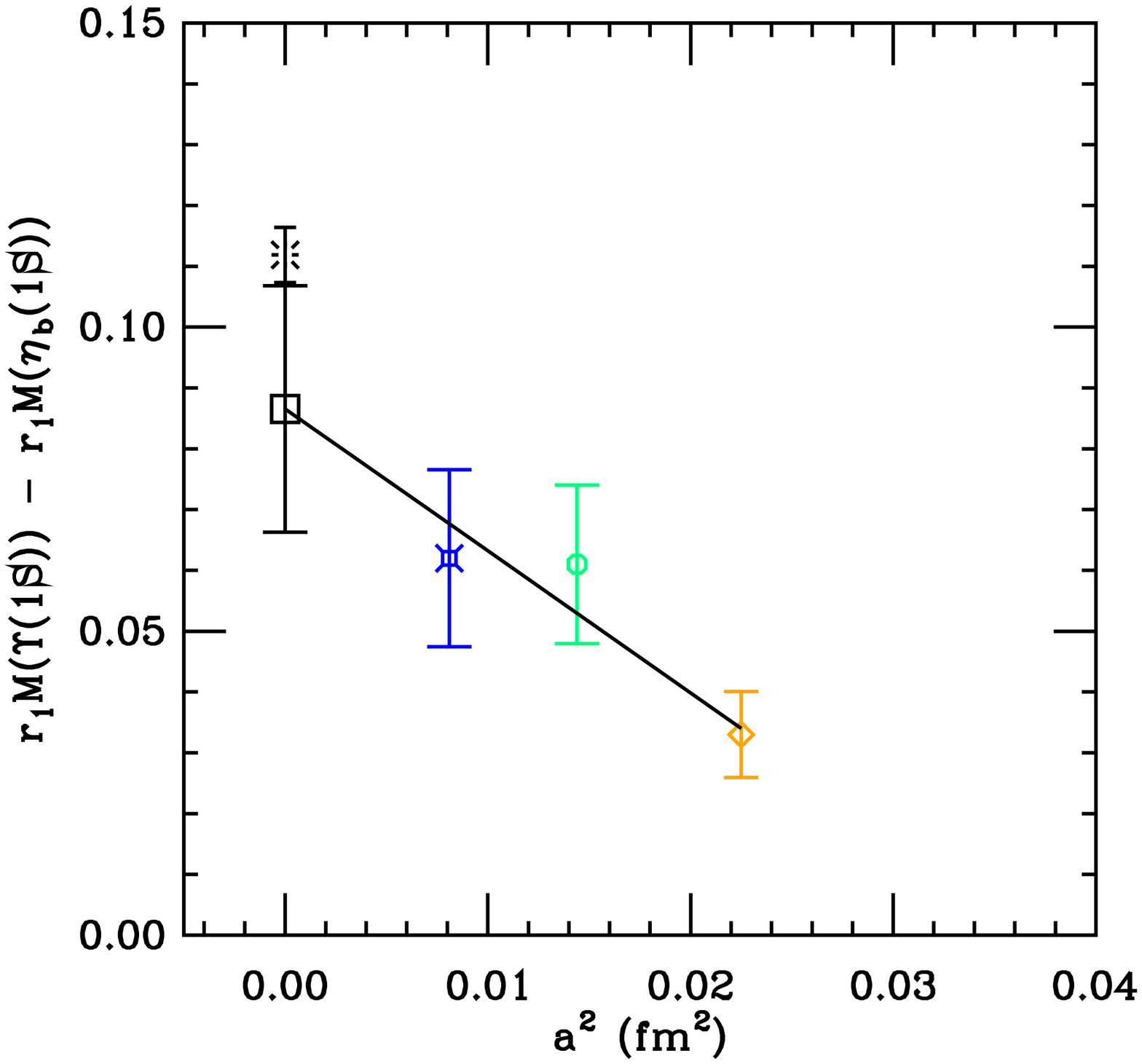}
    \caption{Continuum extrapolations for the $1S$ hyperfine splittings 
        for (a)~charmonium, (b)~bottomonium.
        The symbols and colors of the data points are the same as 
        throughout the paper.
        Here the error bars on the data points include our estimates 
        for the $\kappa$-tuning systematic error.
        The plotted experimental $\eta_b$ mass comes from the average
        of recent measurements~\cite{:2008vj,Aubert:2009pz,Bonvicini:2009hs},
        as discussed in the text.} 
    \label{fig:hfs_vs_a2}
\end{figure*}


Next let us examine the $\overline{2S}$\/-$\overline{1S}$ splitting.
We fit a correlator matrix constructed from two interpolating 
operators, local and smeared, to three or more states 
(\emph{i.e.}, two or more excited states).
The error we assign to the mass determination estimates the 
uncertainties in our method.
The results for charmonium as a function of $(r_1M_\pi)^2$
are shown in Fig.~\ref{fig:2s-1s}a.  
The lattice data appear to lie significantly above the
experimental value at the smaller lattice spacings.
The individual $2S$ levels show the same
trends we observe in the spin-averaged level.  In
Fig.~\ref{fig:cc2s-1s} we plot separately the
$\eta_c(2S)$-$\overline{1S}$ and $\psi(2S)$-$\overline{1S}$.  We see
that both $\eta_c(2S)$ and $\psi(2S)$ are responsible for the behavior
seen in Fig.~\ref{fig:2s-1s}a, the latter especially so.  
The results for bottomonium (Fig.~\ref{fig:2s-1s}b) are more
satisfactory.

We suggest two possible reasons for the behavior of the charmonium 
$\overline{2S}$\/-$\overline{1S}$ splitting results.
First, the $2S$ are the only excited states in this study.
Excited states are more difficult than ground states to determine 
accurately.
With only two operators, our fits are less reliable, even though our 
fit model has at least three states.
Second, the fit procedure does not take into account adequately the 
possible contribution of multiple open charm levels.
For example, we have not used a two-body operator in the matrix 
correlator.
With unphysically large quark masses, the open charm levels are 
unphysically high.
As the sea quark mass is decreased, they come down.
Moreover, the box size of our lattices at the lightest sea 
quark mass is larger, which decreases the discrete level spacing of the 
would-be open-charm continuum.
The dotted line in Fig.~\ref{fig:2s-1s}a shows the location of the 
physical open charm threshold.
It is dangerously close to the physical $2S$ levels, especially the 
$\psi(2S)$.
Thus it is conceivable that nearby multiple open charm levels are being 
confused with the $2S$ and artificially raise its fitted mass.
This explanation is consistent with the observed gradual rise of this 
level in the fine ensembles with decreasing light quark mass but not 
with the trends seen in the coarse and medium-coarse ensembles.

For bottomonium in Fig.~\ref{fig:2s-1s}b, the open bottom threshold is 
safely distant (off scale in this plot), so we do not expect a similar 
confusion in this channel.
Figure~\ref{fig:bb2s-1s} shows the individual $2S$ bottomonium levels 
separately.
There is no comparison for the first excited pseudoscalar state
$\eta_b(2S)$-$\overline{1S}$, because the state has not yet been 
observed~\cite{Amsler:2008zz}, although the extrapolated values appear 
to approach a consistent continuum limit.
The first excited vector state splitting $\Upsilon(2S)$-$\overline{1S}$ 
is given in Fig.~\ref{fig:bb2s-1s}b.
The chirally extrapolated values monotonically approach the experimental 
value and for the fine ensembles our splitting agrees with the experiment.

\begin{figure*}
\centering
    (a)~\includegraphics*[width=8cm]{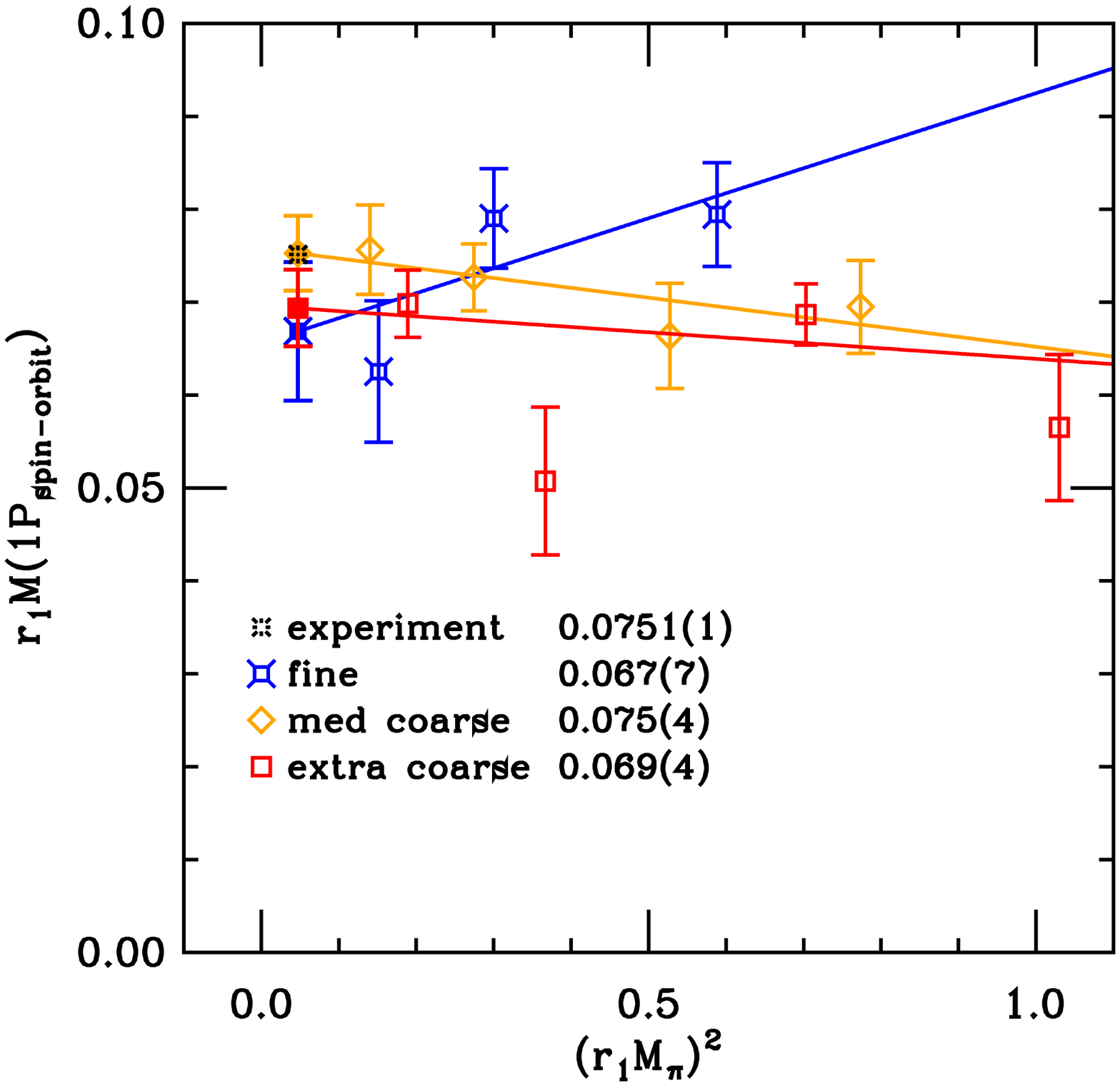}\hfill
    (b)~\includegraphics*[width=8cm]{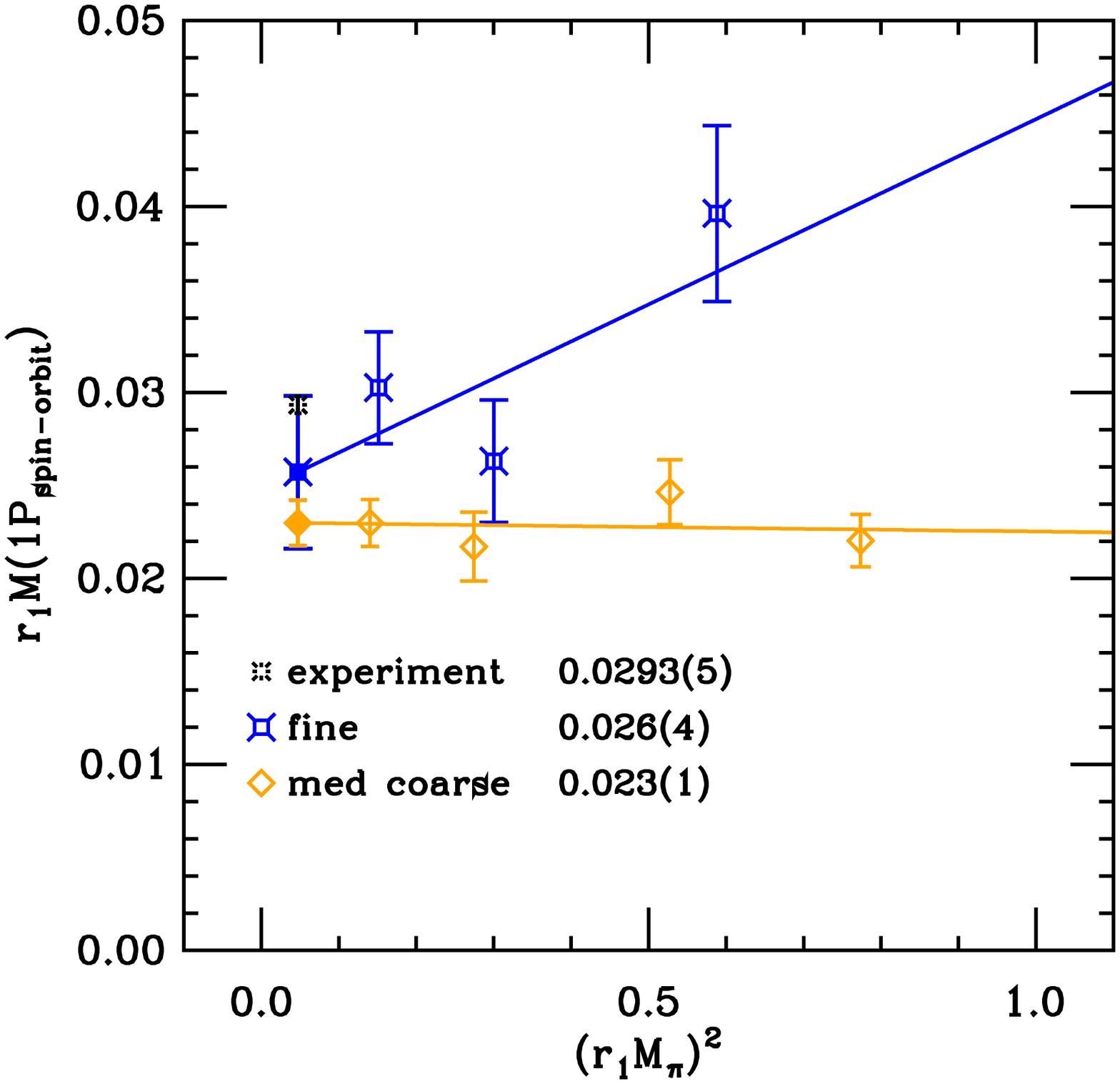}
    \caption{Spin-orbit splittings in $1P$ levels, with 
        $M(1P_{\rm spin-orbit})$ defined in Eq.~(\ref{eq:spin-orbit}), 
        for (a)~charmonium and (b)~bottomonium.}
    \label{fig:chi-spin-orbit}
\end{figure*}
\begin{figure*}
\centering
    (a)~\includegraphics*[width=8cm]{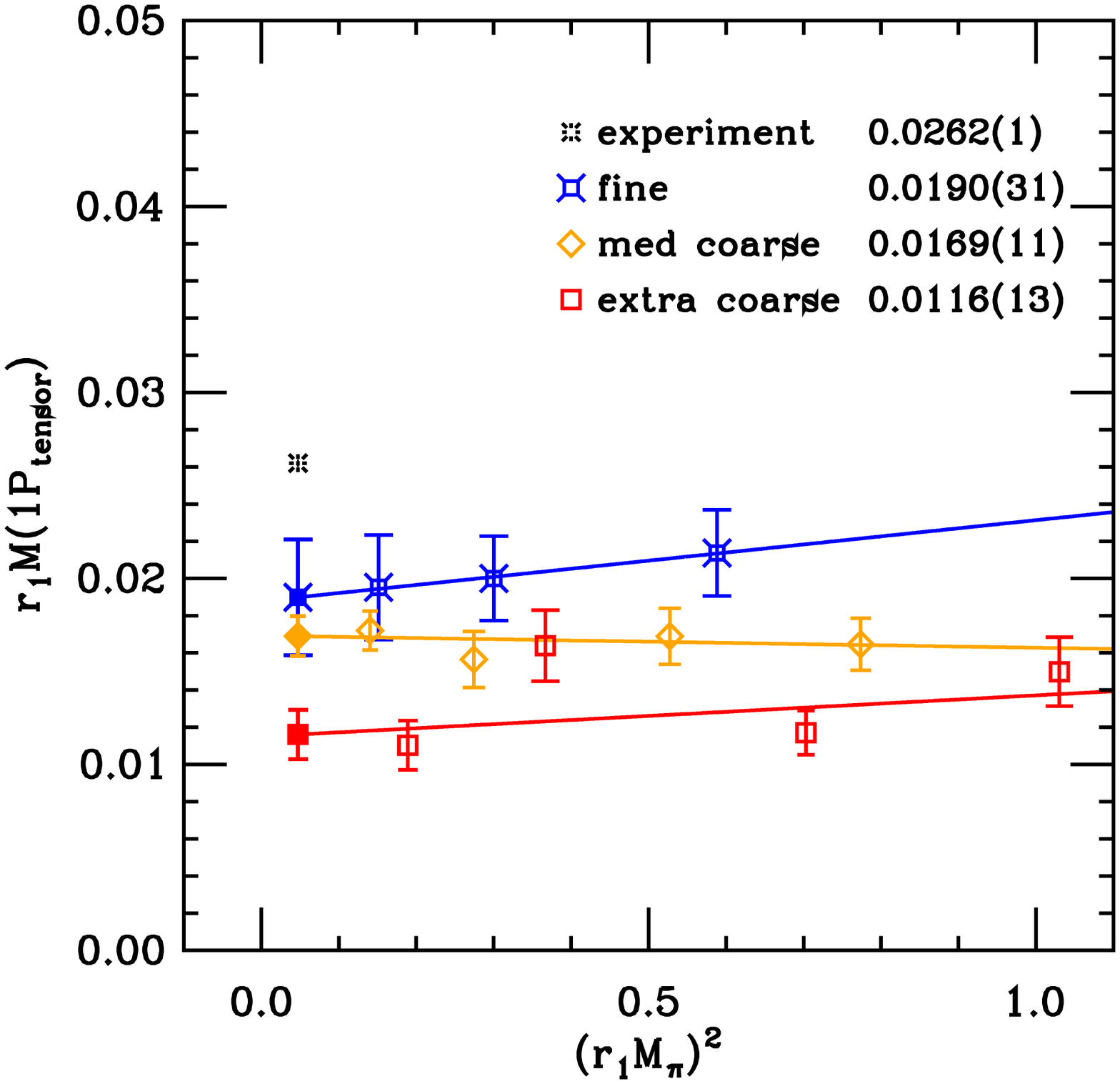}\hfill
    (b)~\includegraphics*[width=8cm]{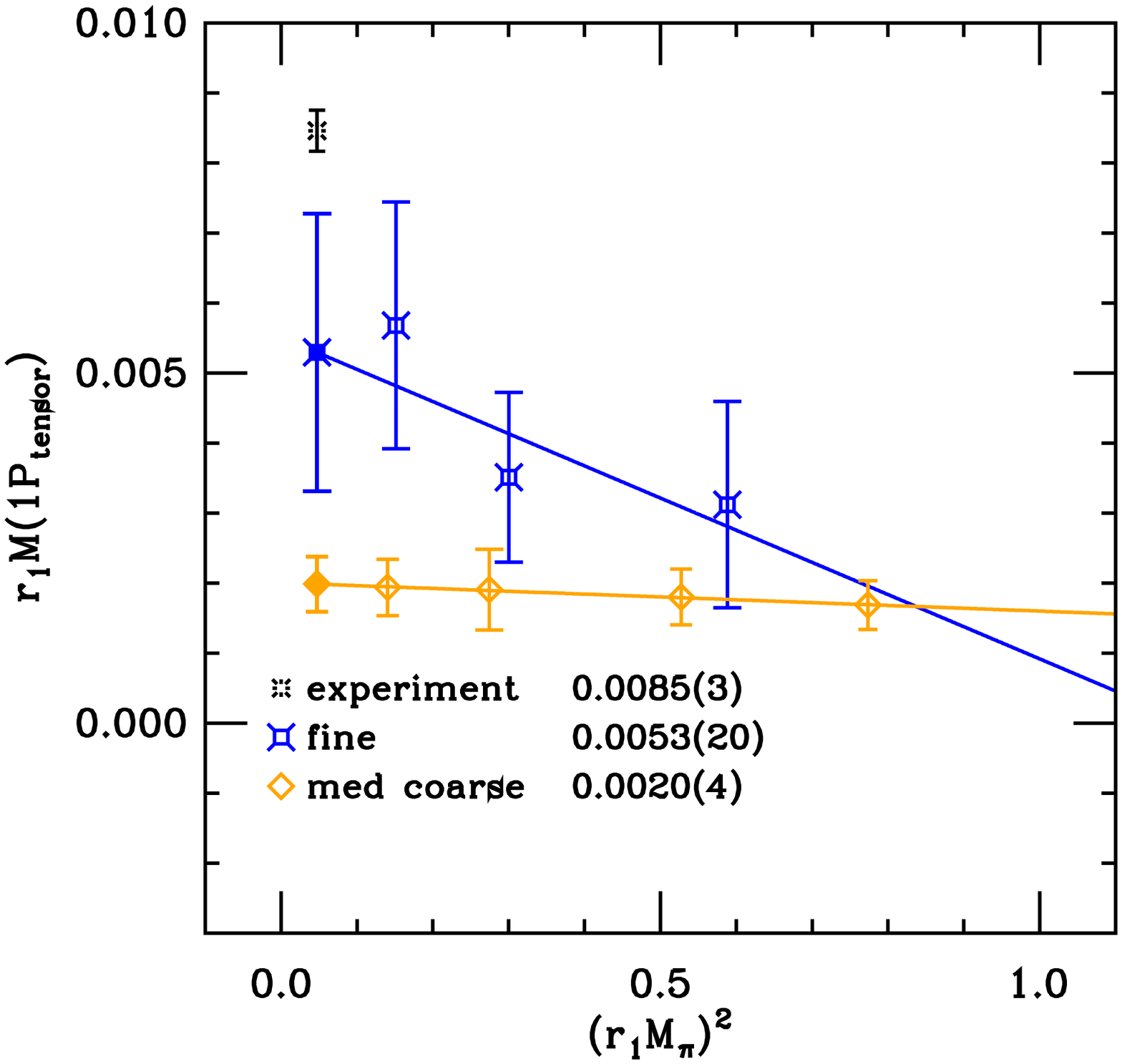}
    \caption{Tensor splittings in $1P$ levels, with 
        $M(1P_{\rm tensor})$ defined in Eq.~(\ref{eq:tensor}), 
        for (a)~charmonium and (b)~bottomonium.}
    \label{fig:chi-tensor}
\end{figure*}
\begin{figure*}
\centering
    (a)~\includegraphics*[width=8cm]{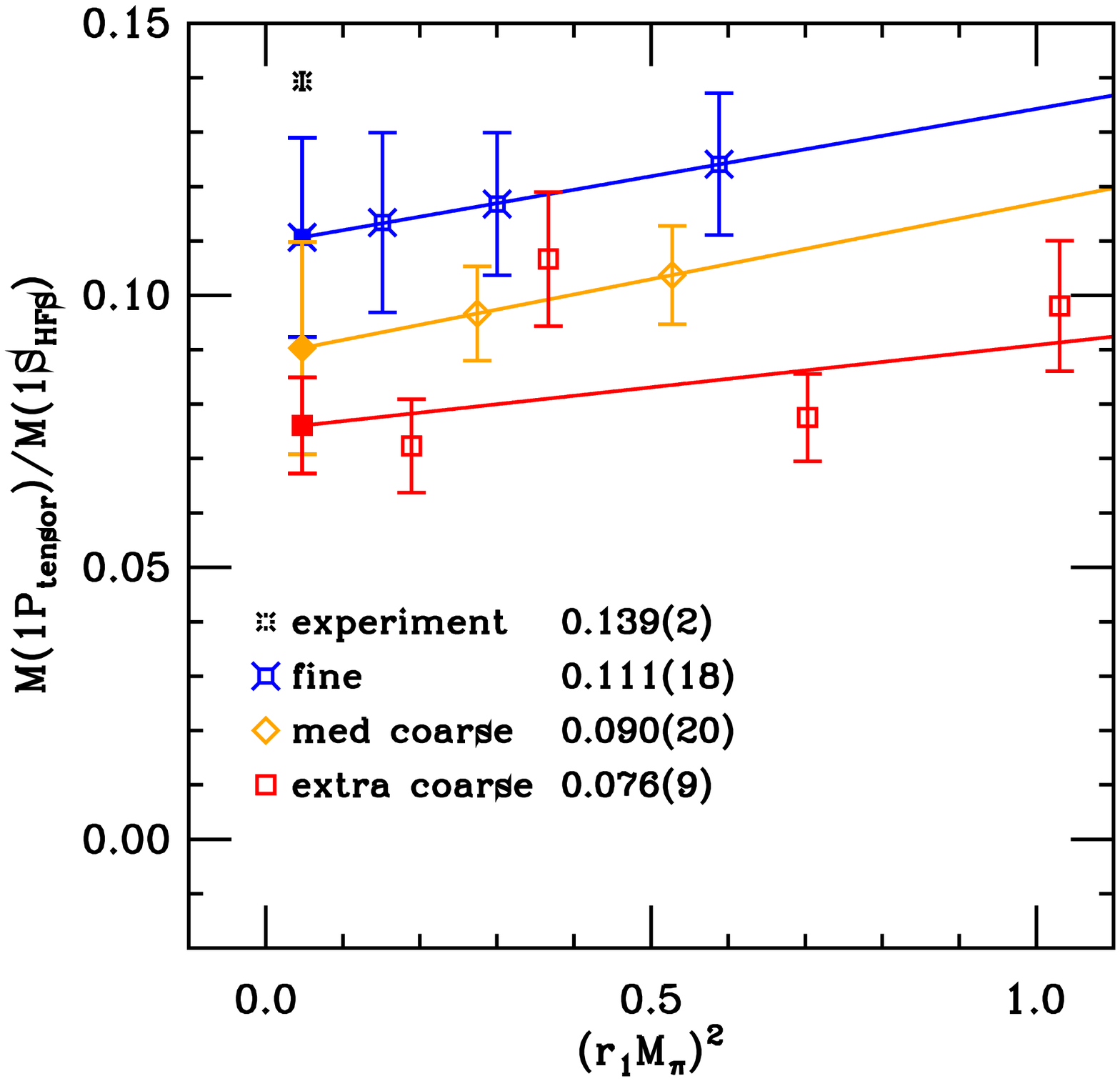}\hfill
    (b)~\includegraphics*[width=8cm]{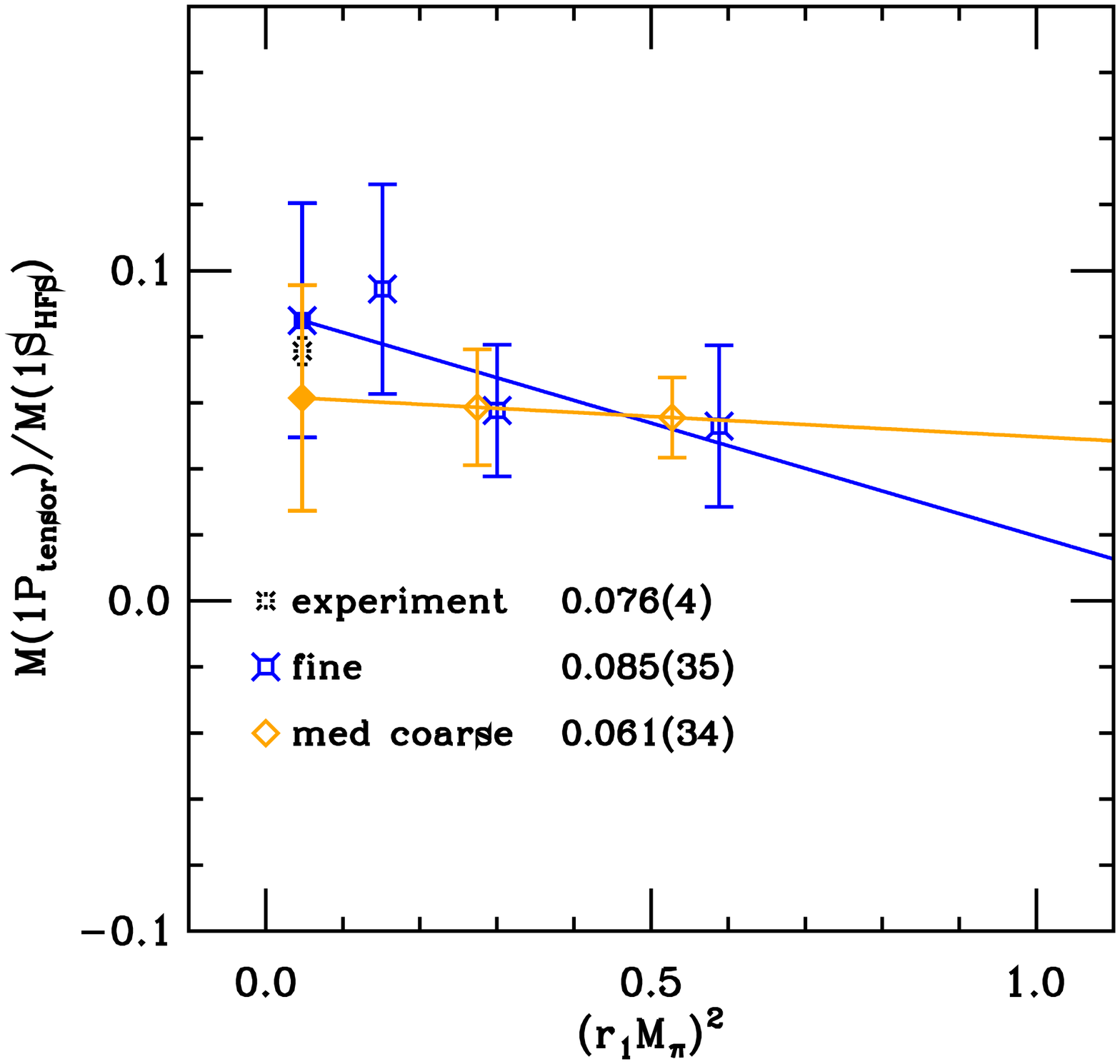}
    \caption{Ratio of the $1P$ tensor and $1S$ hyperfine splittings,
        for (a)~charmonium and (b)~bottomonium.}
    \label{fig:chi-rat}
\end{figure*}

\subsection{Hyperfine splittings}
\label{sec:hfs}


Now let us turn to the hyperfine structure.
Our results for the hyperfine splitting in charmonium and bottomonium 
are presented in Figs.~\ref{fig:cc_hyp}~and~\ref{fig:bb_hyp}. 
For the $1S$ levels and for $2S$ bottomonium, there is little 
dependence on the sea quark mass.
To assess the approach to the continuum limit one must bear in mind 
that the errors in Figs.~\ref{fig:cc_hyp}~and~\ref{fig:bb_hyp} are 
statistical only, and the systematic error from $\kappa$-tuning must 
also be taken into account.
We thus take the values at the physical pion mass, apply the 
$\kappa$-tuning error and plot these data \emph{vs.}~$a^2$,
as shown in Fig.~\ref{fig:hfs_vs_a2}.
Both data sets are consistently linear in~$a^2$,
so we carry out such an extrapolation.
The extrapolated values in units of $r_1$ are $0.187(12)$ for 
charmonium, with $\chi^2/\mathrm{dof}=1.9/2$, and $0.087(20)$ for 
bottomonium, with $\chi^2/\mathrm{dof}=0.55/1$.
One can see, from comparing Fig.~\ref{fig:hfs_vs_a2} with 
Fig.~\ref{fig:cc_hyp}~and~\ref{fig:bb_hyp}, that the $\kappa$-tuning 
uncertainties inherited from the heavy-strange kinetic mass are larger 
than the statistical uncertainties of the quarkonium rest-mass 
splittings.

In physical units these extrapolated results are 
$M_{J/\psi(1S)}-M_{\eta_c(1S)}=116.0\pm 7.4^{+2.6}_{-0.0}$~MeV 
and $M_{\Upsilon(1S)}-M_{\eta_b(1S)}=54.0\pm 12.4^{+1.2}_{-0.0}$~MeV,
where the second error comes from converting from $r_1$ units to~MeV.
For charmonium the average of experimental measurements is 
$116.4\pm 1.2$~MeV \cite{Amsler:2008zz}, so our result is perfectly 
consistent.
For bottomonium, the experimental measurements are
$71.4^{+2.3}_{-3.1}\pm2.7$~MeV \cite{:2008vj},
$66.1^{+4.9}_{-4.8}\pm2.0$~MeV \cite{Aubert:2009pz}, and
$68.5\pm 6.6       \pm2.0$~MeV \cite{Bonvicini:2009hs};
symmetrizing the error bars and taking a weighted average,
we find $M_{\Upsilon(1S)}-M_{\eta_b(1S)}=69.4\pm2.8$~MeV.
Our hyperfine splitting thus falls $1.2\sigma$ short.
Note that with lattice NRQCD, the HPQCD Collaboration finds
$M_{\Upsilon(1S)}-M_{\eta_b(1S)}=61\pm4\pm13$~MeV \cite{Gray:2005ur}, 
which agrees with the recent experimental measurements,
yet also with our result.

The errors on the final $1S$~hyperfine splittings quoted here encompass 
statistics (as amplified by extrapolations), $\kappa$~tuning, and $r_1$.
In addition, the coupling $c_B$ has been adjusted only at the 
tree-level, introducing an error of ${\rm O}(\alpha_sa)$ that our 
continuum extrapolation would not eliminate.
A preliminary result for the one-loop correction to $c_B$ is 
available~\cite{Nobes:2005dz}, suggesting that a very small 
correction is needed beyond the tadpole improvement of 
Eq.~(\ref{eq:cBu0}), when $u_0$ is set from the Landau link.

\begin{figure*}
\centering
    (a)~\includegraphics*[width=8cm]{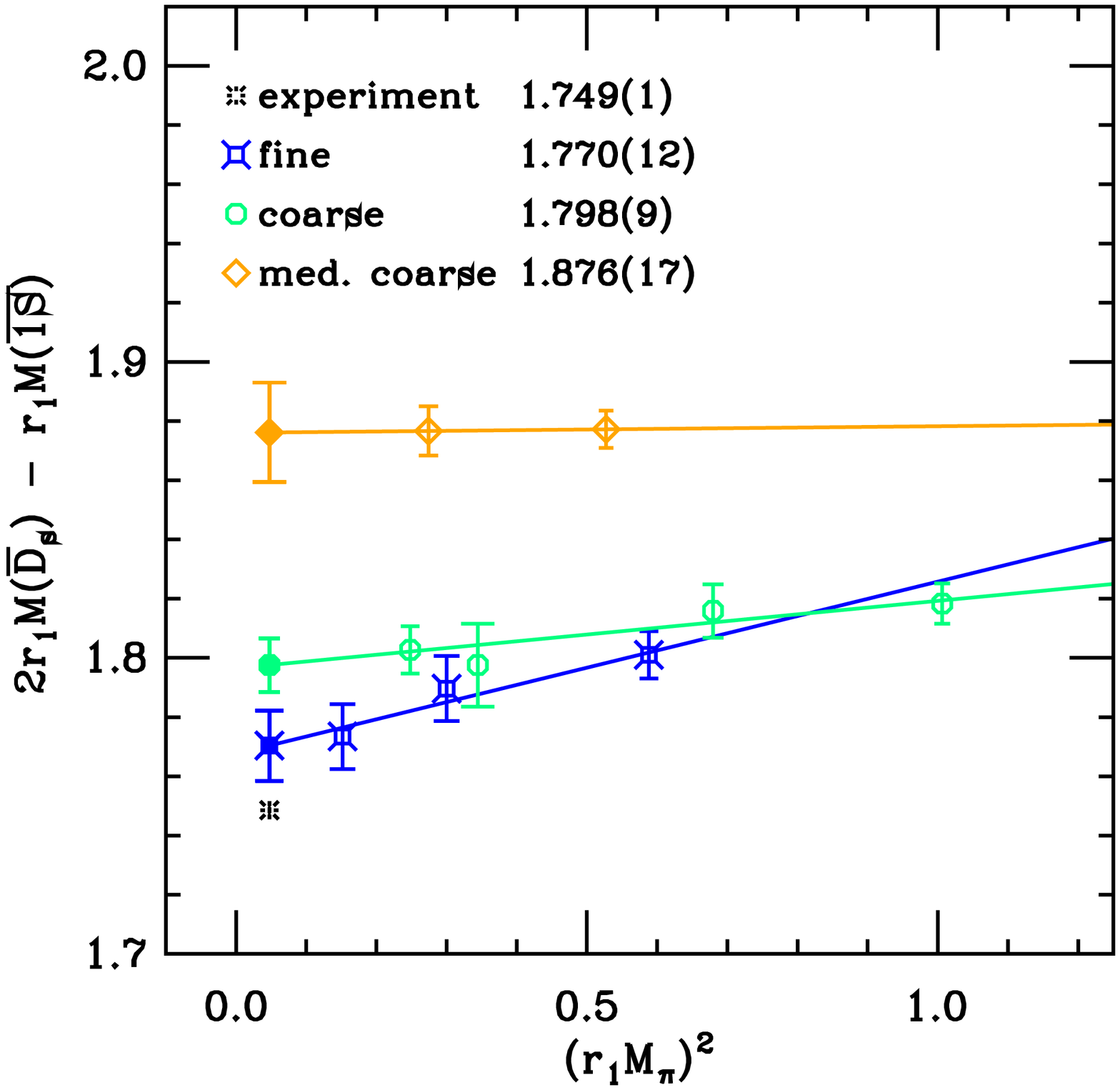}\hfill
    (b)~\includegraphics*[width=8cm]{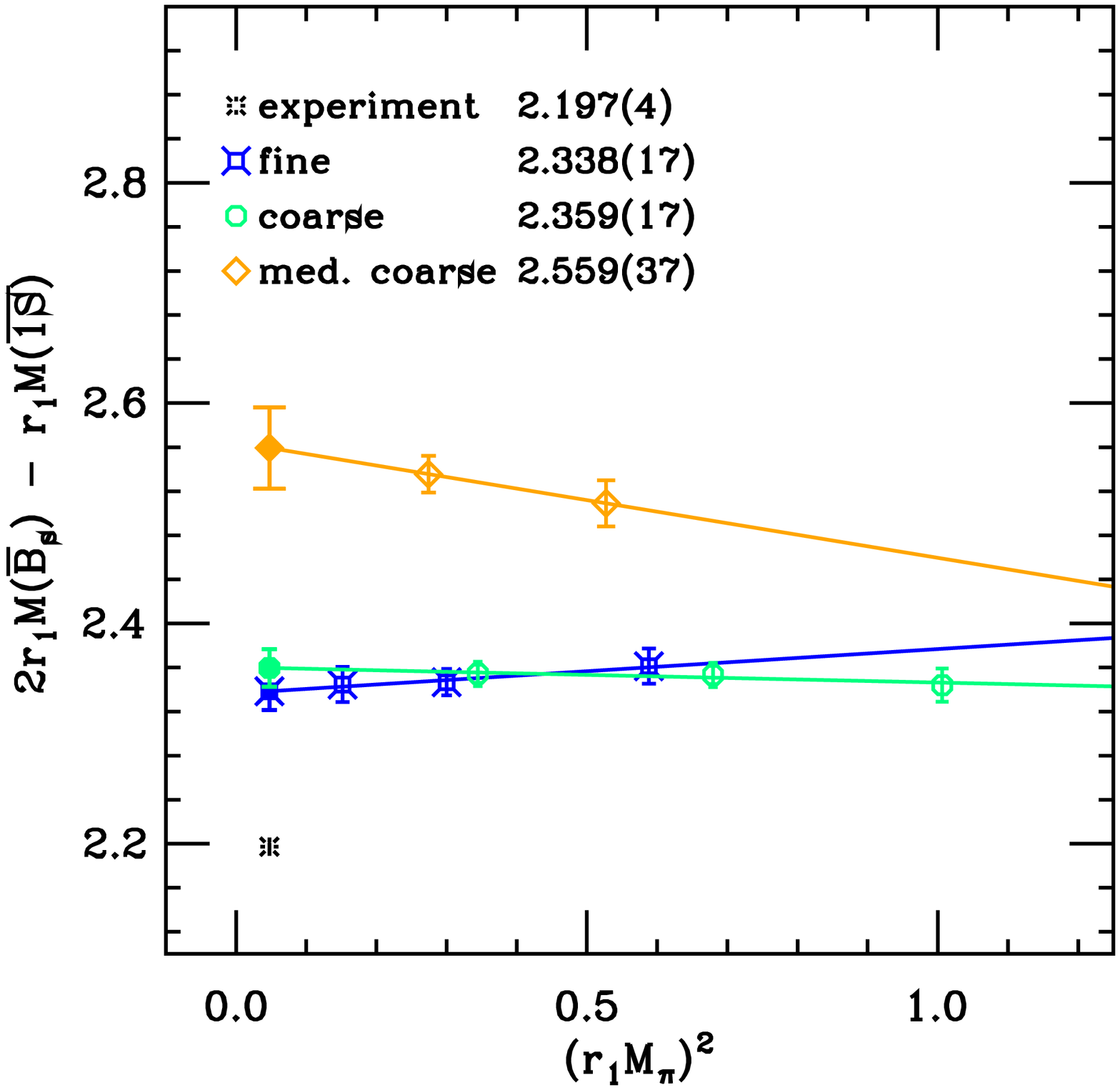}
    \caption{Quarkonium--heavy-light splittings 
        (a)~$2M(\overline{D_s})-M(\overline{1S})$ and 
        (b)~$2M(\overline{B_s})-M(\overline{1S})$.}
    \label{fig:1S-2DB}
\end{figure*}
\begin{figure*}
\centering
    (a)~\includegraphics*[width=8cm]{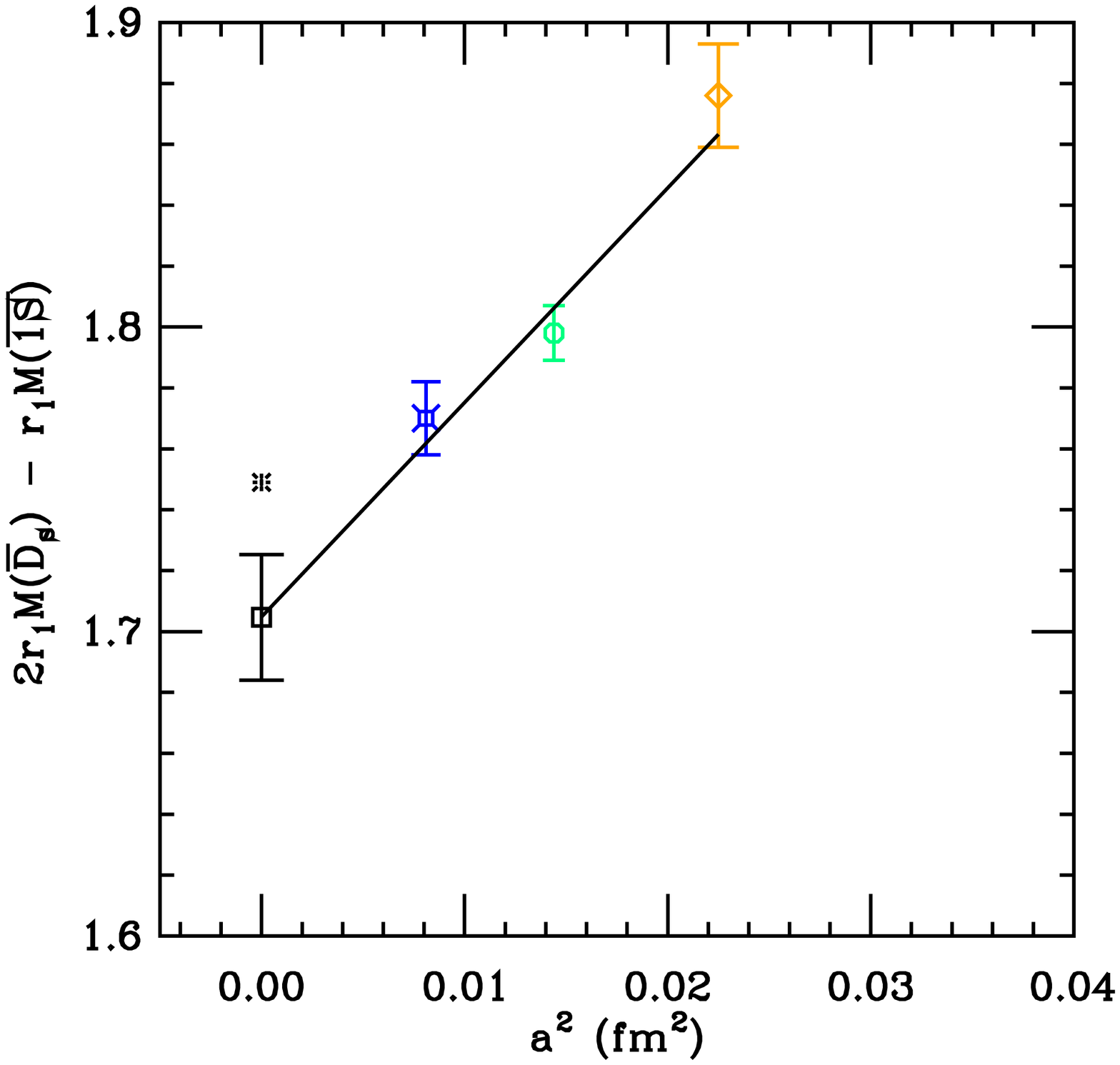}\hfill
    (b)~\includegraphics*[width=8cm]{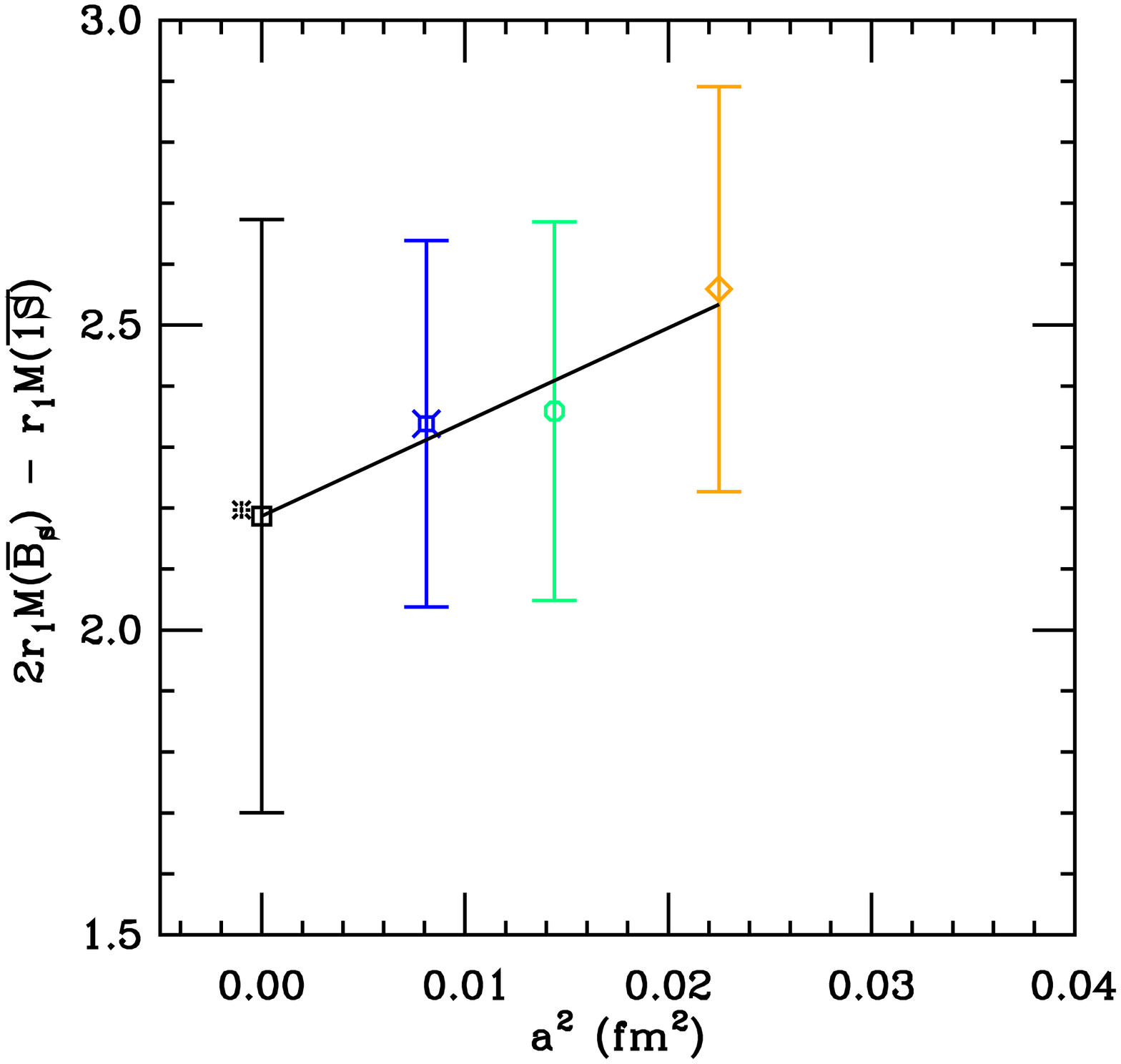}
    \caption{Continuum extrapolations of 
        (a)~$2M(\overline{D_s})-M(\overline{1S})$ and 
        (b)~$2M(\overline{B_s})-M(\overline{1S})$.}
    \label{fig:1S-2DBcont}
\end{figure*}
%


The $2S$ hyperfine splittings for both charmonium and bottomonium
are shown in Figs.~\ref{fig:cc_hyp}b~and~\ref{fig:bb_hyp}b.
Unfortunately, these results are not very useful.
Although the charmonium splitting agrees, within large errors, with 
experiment, one should bear in mind the issue of threshold effects 
surrounding our determination of the $\psi(2S)$ mass, discussed above.
The bottomonium splitting does not suffer from this problem, but the 
statistical and fitting errors are still too large to make a prediction 
of the as yet unobserved $\eta_b(2S)$ mass.

\begin{figure*}
\centering
    \includegraphics*[width=8cm]{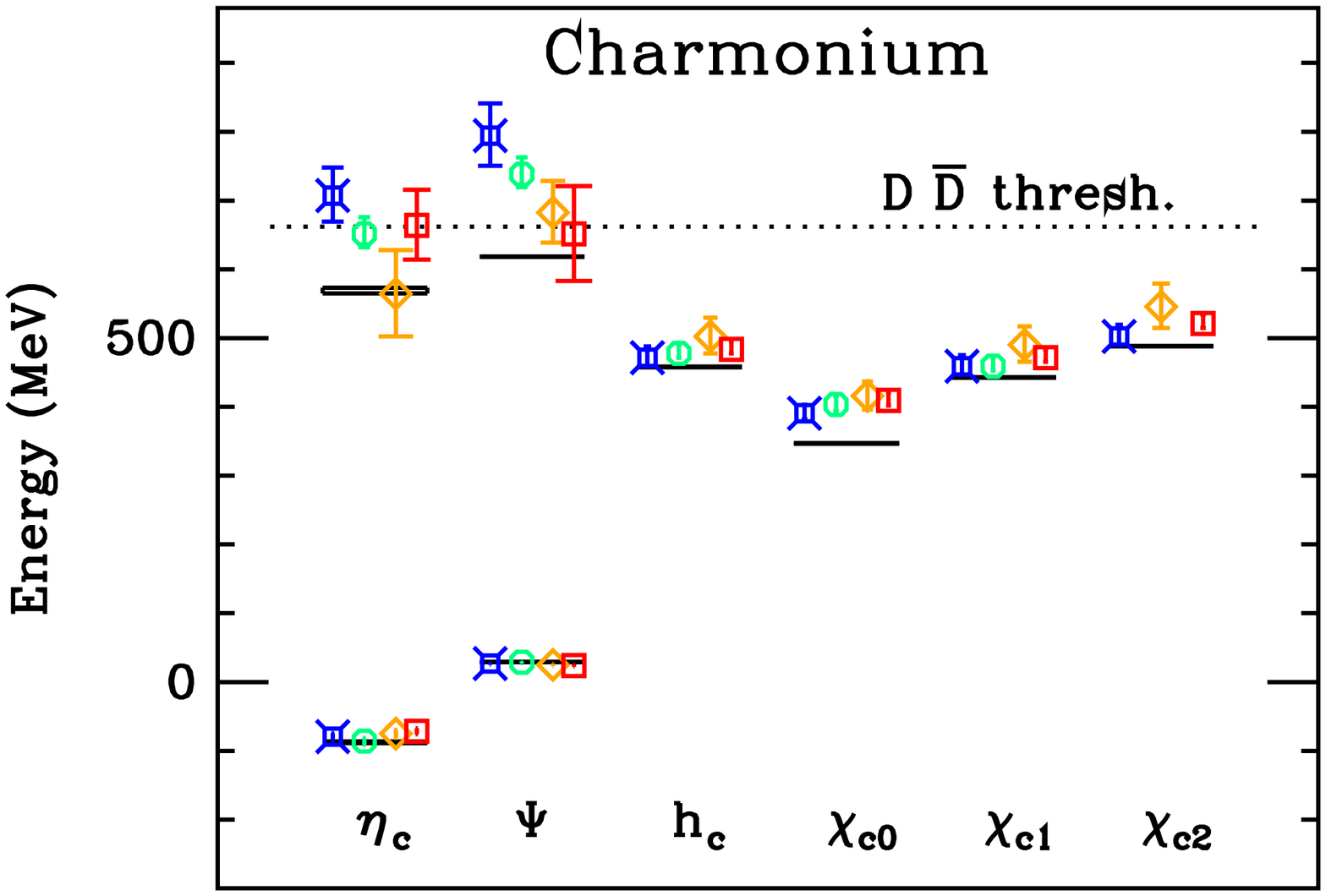}\hfill
    \includegraphics*[width=8cm]{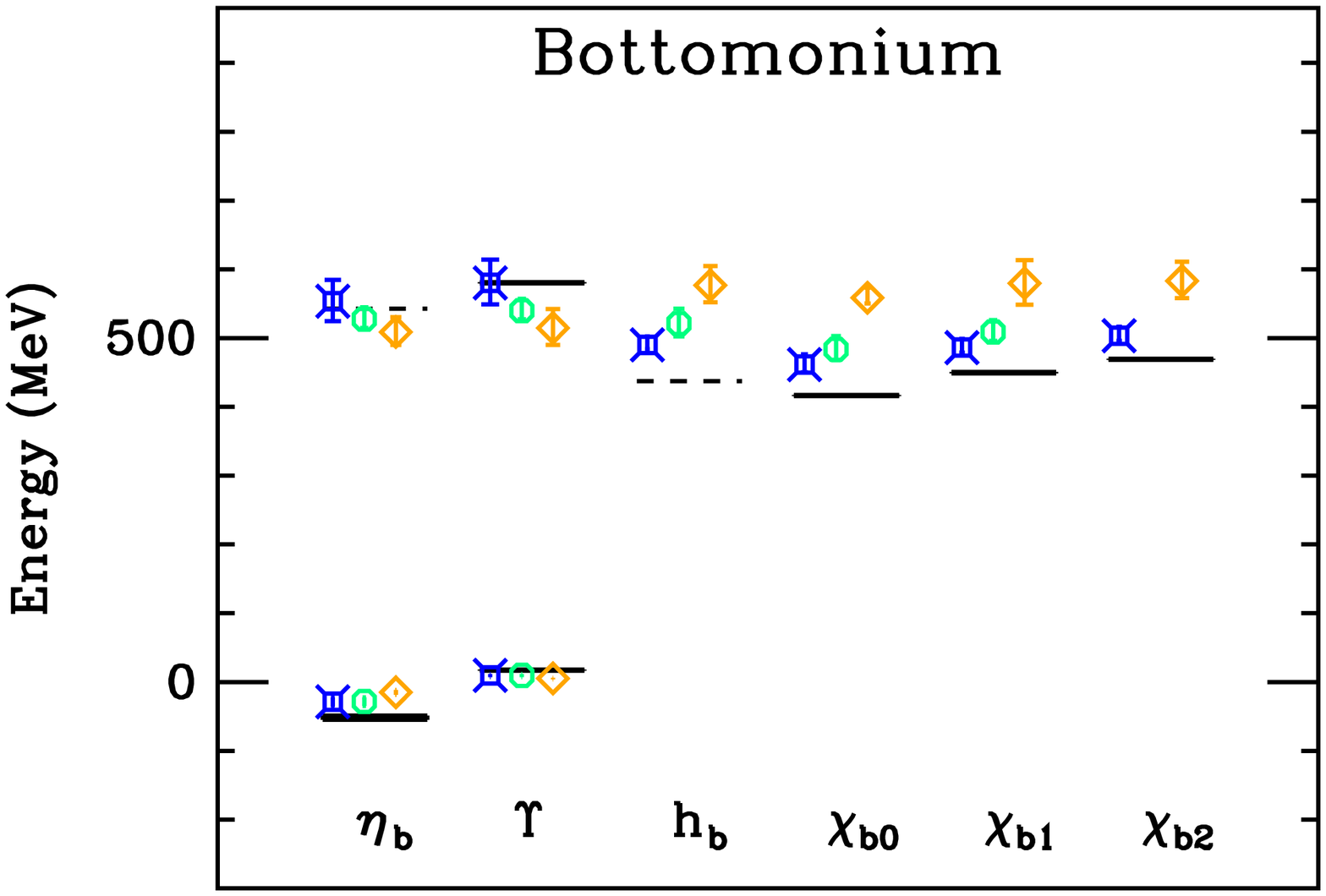}
    \caption{Quarkonium spectrum as splittings from the 
        $\overline{1S}$~level for $\bar{c}c$ (left) and $\bar{b}b$ 
        (right).
        The fine-ensemble results are in blue fancy squares, the coarse
        in green circles, the medium-coarse in orange diamonds and the 
        extra-coarse in red squares.
        Solid lines show the experimental values,
        and dashed lines estimates from potential models.
        The dotted line in the left panel indicates the physical 
        open-charm threshold.
        The error on the data points combines statistical, 
        $\kappa$-tuning, and $r_1$ uncertainties.}
    \label{fig:qq-all}
\end{figure*}

\subsection{$P$-state splittings}


We now turn to splittings between the $1^3\!P_J$ levels, which stem from 
two contributions~\cite{Eichten:1980mw}.
As discussed above, one comes from exchanging a Coulomb gluon between a 
spin-orbit term,  
$\bar{h}^{(\pm)}i\bm{\sigma}\cdot(\bm{D}\times\bm{E})h^{(\pm)}$ in 
Eq.~(\ref{eq:Leff4}), and the static potential, 
$\bar{h}^{(\mp)}A_4h^{(\mp)}$.
The other comes from exchanging a transverse gluon between the 
chromomagnetic terms, $\bar{h}^{(+)}i\bm{\sigma}\cdot\bm{B}h^{(+)}$ 
and $\bar{h}^{(-)}i\bm{\sigma}\cdot\bm{B}h^{(-)}$.
These two contributions can be separated by forming the 
combinations in Eqs.~(\ref{eq:spin-orbit}) and~(\ref{eq:tensor})
\cite{Peskin:1983up}.

The spin-orbit splittings~$M(1P_{\textrm{spin-orbit}})$ are shown in 
Fig.~\ref{fig:chi-spin-orbit} for charmonium and bottomonium.
They exhibit a small lattice-spacing dependence and agree well with 
experiment, indicating that the chromoelectric interactions and, hence, 
$c_E$ are adjusted accurately enough.
The tensor splittings~$M(1P_{\rm tensor})$ are shown in 
Fig.~\ref{fig:chi-tensor} for charmonium and bottomonium.
These chromomagnetic effects seem to approach the experimental value as 
$a$ decreases.
Since the tensor and the spin-spin potential components both measure 
the effects of the chromomagnetic interaction, we plot the ratio of the 
$1P$ tensor splitting to the $1S$ hyperfine splitting in 
Fig.~\ref{fig:chi-rat}.
The coefficient~$m_B^{-1}$ should drop out from the ratio and if there 
are no effects from higher-dimension operators, the ratio should be a 
constant which agrees with the continuum limit. 
If a higher-dimension operator has a significant contribution, then the 
ratio need not agree any better than the splittings themselves.
The charmonium case, Fig.~\ref{fig:chi-rat}a, seems to suggest that the 
higher-dimension operator matters, the bottomonium case, 
Fig.~\ref{fig:chi-rat}b, seems to suggest it does not.
This outcome is plausible, because the $v^2$ suppression of the 
higher-dimension operator is 10\% in bottomonium, but only 30\% in 
charmonium [cf.\ Eqs.~(\ref{eq:ccvp}) and~(\ref{eq:bbvp})].

\subsection{Quarkonium \emph{vs.}\ heavy-strange mesons}
\label{sec:QQsQ}

Unlike other approaches to heavy quarks, lattice QCD is supposed to 
treat heavy-light mesons and quarkonium on the same footing.
If we form the splitting
\begin{equation}
    2M(\overline{D_s}) - M(\overline{1S})
    \label{eq:QQ-qQ}
\end{equation}
the rest mass drops out, leaving a pure QCD quantity.
Here $M(\overline{D_s})$ denotes the spin average of $D_s$ and $D^*_s$ 
masses.
This mass difference is interesting from the point of view of the 
discretization effects, which should contribute less to the 
$\overline{D_s}$ and $\overline{B_s}$ than to the charmonium and 
bottomonium $\overline{1S}$ states.
We show this splitting (also for the bottom-quark sector) combining our 
quarkonium rest masses with the Fermilab-MILC heavy-strange rest 
masses~\cite{Freeland:2009} in Fig.~\ref{fig:1S-2DB}.
The correlation in the error is treated correctly with the bootstrap 
method, and, as elsewhere in this paper, the bootstrap errors are 
symmetrized.
Clearly, discretization effects are important at nonzero~$a$.

In Fig.~\ref{fig:1S-2DBcont}, we incorporate the $\kappa$-tuning errors 
and show the $a$ dependence of the above splittings.
Carrying out an exptrapolation linear in $a^2$, which is empirically 
suitable, we find
$r_1[2M(\overline{D_s})-M(\overline{1S})]=1.705\pm 0.021$ and
$r_1[2M(\overline{B_s})-M(\overline{1S})]=2.19\pm 0.49$;
these correspond to
$2M(\overline{D_s})-M(\overline{1S})=1058\pm 13^{+24}_{-0}$~MeV and
$2M(\overline{B_s})-M(\overline{1S})=1359\pm 304^{+31}_{-0}$~MeV,
with the uncertainty in $r_1$ yielding the second error bar.
The bottomonium extrapolation agrees with the experimental value, but the 
combined statistical and $\kappa$-tuning errors are quite large.
The charmonium extrapolation is $1\sigma$ shy of the experimental value.
Given the empirical nature of our continuum extrapolation, this is 
completely satisfactory.

\subsection{Summary of spectrum results}
\label{sec:sum}

To summarize our results, Fig.~\ref{fig:qq-all} shows the charmonium and 
the bottomonium spectra as splittings from the $\overline{1S}$ level and 
compares them to the experimental results.
We have plotted the chirally extrapolated values at each lattice 
spacing and included statistical, $\kappa$-tuning, and $r_1$ 
uncertainties. 
Solid lines show the experimental values, where they are known, and
dashed lines show estimates from potential 
models~\cite{Buchmuller:1980su} in other cases.

%
%
\begin{table*}
    \centering
    \caption{Continuum extrapolations of splittings in charmonium and 
        bottomonium in MeV. 
        The first error comes from statistics and accumulated 
        extrapolation systematics; the second comes from the 
        uncertainty in scale setting with $r_1=0.318^{+0.000}_{-0.007}$~fm.}
    \label{tab:summary}
    \begin{tabular*}{0.7\textwidth}{c*{4}{@{\extracolsep{\fill}}l}}
    \hline\hline
    Splitting &\multicolumn{2}{c}{Charmonium}&\multicolumn{2}{c}{Bottomonium}\\
              &   This work & Experiment & This work & Experiment \\
     \hline
     $\overline{1P}$-$\overline{1S}$ & $473 \pm 12^{+10}_{-0}$ & $457.5 \pm 0.3$ &
        $446 \pm 18^{+10}_{-0}$ & $456.9 \pm 0.8$ \\
     ${}^1\!P_1$-$\overline{1S}$     & $469 \pm 11^{+10}_{-0}$ & $457.9 \pm 0.4$ &
        $440 \pm 17^{+10}_{-0}$ &  ~~~~~--- \\
     $\overline{2S}$-$\overline{1S}$ & $792 \pm 42^{+17}_{-0}$ & $606 \pm 1$ &
        $599 \pm 36^{+13}_{-0}$ & $(580.3 \pm 0.8)$%
        \footnote{$\Upsilon(2S)$-$\overline{1S}$ instead of $\overline{2S}$-$\overline{1S}$.} \\
     $1^{3}\!S_1$-$1^{1}\!S_0$       & $116.0 \pm 7.4^{+2.6}_{-0}$ & $116.4\pm1.2$ &
        $54.0 \pm 12.4^{+1.2}_{-0}$ & $69.4 \pm 2.8$ \\
     $1P$~tensor                     & $15.0 \pm 2.3^{+0.3}_{-0}$ & $16.25\pm0.07$ &
        $4.5 \pm 2.2^{+0.1}_{-0}$ & $5.25 \pm 0.13$ \\
     $1P$~spin-orbit                 & $43.3 \pm 6.6^{+1.0}_{-0}$ & $46.61 \pm 0.09$ &
        $16.9 \pm 7.0^{+0.4}_{-0}$ & $18.2 \pm 0.2$ \\
     $1S~\bar{s}Q$-$\bar{Q}Q$           & $1058 \pm 13^{+24}_{-0}$ & $1084.8 \pm 0.8$ &
        $1359 \pm 304^{+31}_{-0}$ & $1363.3 \pm 2.2$ \\
    \hline\hline
    \end{tabular*}
\end{table*}

For the splittings discussed above, Table~\ref{tab:summary} shows the 
continuum limit, taken via linear extrapolations in~$a^2$.
One should bear in mind that the NRQCD-based theory of cutoff effects, 
explained in Sec.~\ref{sec:heavy}, anticipates a less trivial 
lattice-spacing dependence.
The linear-in-$a^2$ extrapolations are consistent with the data,
which are not yet sufficient to resolve more complicated functional 
forms.
In Table~\ref{tab:summary} the second (asymmetric) error bar comes from 
the conversion to MeV with 
$r_1=0.318^{+0}_{-0.007}~\text{fm}=1.611^{+0}_{-0.035}~\text{GeV}^{-1}$~%
\cite{Bazavov:2009bb,Bazavov:2009fk,Davies:2009ts}.

The charmonium and bottomonium spectra by and large show good agreement 
with experiment.
The charmonium hyperfine splitting agrees very well;
the bottomonium splitting agrees at $1.2\sigma$.
The tensor and spin-orbit splittings also agree well, for both systems.
The ${}^1\!P_1$-$\overline{1S}$ and $\overline{1P}$-$\overline{1S}$ 
spin-averaged splittings agree at 1.1--1.3$\sigma$ for $\bar{c}c$;
the $\overline{1P}$-$\overline{1S}$ at $0.6\sigma$ for $\bar{b}b$.
As discussed above, the charmonium $2S$ states are too high, because 
our operator basis and statistics proved to be insufficient to 
disentangle the bound states from open-charm threshold effects.
For bottomonium the $\overline{2S}$-$\overline{1S}$ splitting does not 
suffer from threshold effects and agrees well.
When the $r_1$ uncertainty is included, the splitting of quarkonium relative 
to the heavy-strange spectrum, $2M(\overline{D_s})-M(\overline{1S})$ and 
$2M(\overline{B_s})-M(\overline{1S})$, also agrees well with experiment. 

\section{Conclusions}
\label{sec:conc}

Quarkonium properties offer an excellent test of lattice QCD, because 
they are relatively well-understood hadrons, via potential models and 
effective field theories.
This paper attempts a thorough study of the charmonium and bottomonium 
mass splittings, using lattice gauge fields with 2+1 flavors of sea 
quarks.
By using the Fermilab method for heavy quarks, we are able to study 
both systems, as well as heavy-light hadrons, with the same basic 
theoretical tool.
By using the MILC ensembles, we are able to study a wide range of 
lattice spacing, and a wide range of up and down sea-quark masses, 
down to 0.10$m_s$.

Our aim here has been to develop methods and to compare discretization
effects against expectations that are gleaned from an effective theory 
analysis.
An important technical finding for ground $P$ states is that nonrelativistic 
operators are superior to relativistic operators in overlap and, hence, 
statistics.

Our calculations reproduce most features of the mass splittings, to the 
extent expected.
This optmistic conclusion is marred somewhat, because we find that the 
errors from $\kappa$ tuning are significant for spin-dependent 
splittings.
Agreement with experiment is found only when these uncertainties, which 
stem from the heavy-strange kinetic mass, are taken in to account.
In some other cases, such as leptonic decay 
constants for heavy-light mesons~\cite{Aubin:2005ar}, uncertainties in 
$\kappa$ also influence significantly the final error budget.

In the continuation of this project, we hope to improve on the results 
presented here in several ways.
First, the MILC ensembles now contain approximately four times as many
configurations, and they extend to smaller lattice spacings,
$a\approx0.06$~fm and $a\approx0.045$~fm.
The finer lattice will bring charm into the region where
Symanzik-motivated continuum extrapolations are justified and should
bring bottomonium discretization effects under 1\%.
To this end it may also prove worthwhile to incorporate the $p^4$
corrections of the improved Fermilab action~\cite{Oktay:2008ex}.
Higher statistics and twisted-boundary conditions~\cite{Bedaque:2004kc}
should improve the tuning of $\kappa$ and, thus, reduce errors from 
this source as well.

\acknowledgments

Computations for this work were carried out 
on facilities of the USQCD Collaboration, which are funded by the 
Office of Science of the U.S. Department of Energy.
This work was supported in part by the U.S. Department of
Energy under Grants No.~DE-FC02-06ER41446 (T.B., C.D., L.L.),
No.~DE-FG02-91ER40661 (S.G.), No.~DE-FG02-91ER40677 (A.X.K.),
No.~DE-FG02-91ER40628 (E.D.F.); by the National Science Foundation 
under Grants No.~PHY-0555243, No.~PHY-0757333, 
No.~PHY-0703296 (T.B., C.D., L.L.), and No.~PHY-0555235 (E.D.F.);
and by the M. Hildred Blewett Scholarship of the American Physical 
Society (E.D.F.).
Fermilab is operated by Fermi Research Alliance, LLC, under Contract 
No.~DE-AC02-07CH11359 with the U.S. Department of Energy.

\end{document}